%% file: manuscript.tex
\documentclass[twocolumn, times, trackchanges]{aastex63_bren}
\usepackage{CJK}
\usepackage{booktabs}
\usepackage{xcolor, soul}
\usepackage{amsmath, amssymb, bm}
\usepackage{graphicx}
\usepackage{dsfont}
\graphicspath{{figures/}}
\graphicspath{{./}{figures/}}

\usepackage{float}

\submitjournal{ApJ}

\shorttitle{Cycle-StarNet}
\shortauthors{O'Briain et al.}

\input{macros}

\begin{document}
\begin{CJK*}{UTF8}{gbsn}

\title{Cycle-StarNet: Bridging the gap between theory and data \\ by leveraging large datasets}

\correspondingauthor{Teaghan B. O'Briain}
\email{obriaint@uvic.ca}

\author{Teaghan O'Briain}
\affiliation{Department of Physics and Astronomy, University of Victoria,Victoria, BC, V8W 3P2}

\author{Yuan-Sen Ting (丁源森)}
\affiliation{Institute for Advanced Study, Princeton, NJ 08540, USA}
\affiliation{Department of Astrophysical Sciences, Princeton University, Princeton, NJ 08540, USA}
\affiliation{Observatories of the Carnegie Institution of Washington, 813 Santa Barbara Street, Pasadena, CA 91101, USA}
\affiliation{Research School of Astronomy \& Astrophysics, Australian National University, Cotter Rd., Weston, ACT 2611, Australia}

\author{S\'ebastien Fabbro}
\affiliation{Department of Physics and Astronomy, University of Victoria,Victoria, BC, V8W 3P2}
\affiliation{National Research Council Herzberg Astronomy \& Astrophysics, 4071 West Saanich Road, Victoria, BC}

\author{Kwang M. Yi}
\affiliation{Department of Computer Science, University of British Columbia, Vancouver, BC, V6T 1Z4}

\author{Kim Venn}
\affiliation{Department of Physics and Astronomy, University of Victoria,Victoria, BC, V8W 3P2}

\author{Spencer Bialek}
\affiliation{Department of Physics and Astronomy, University of Victoria,Victoria, BC, V8W 3P2}

\begin{abstract}
The advancements in stellar spectroscopy data acquisition have made it necessary to accomplish similar improvements in efficient data analysis techniques. Current automated methods for analyzing spectra are either (a) data-driven, which requires prior knowledge of stellar parameters and elemental abundances, or (b) based on theoretical synthetic models that are susceptible to the gap between theory and practice. In this study, we present a \textit{hybrid} generative domain adaptation method that turns simulated stellar spectra into realistic spectra by applying unsupervised learning to large spectroscopic surveys. \YST{We apply our technique to the APOGEE H-band spectra at $R=22,500$ and the Kurucz synthetic models.} As a proof of concept, two case studies are presented. The first of which is \obriain{the calibration of synthetic data to become consistent with observations.} To accomplish this, synthetic models are morphed into spectra that resemble observations, thereby reducing the gap between theory and observations. \obriain{Fitting the observed spectra shows an improved average reduced $\chi_R^2$ from 1.97 to 1.22, along with a reduced mean residual from 0.16 to -0.01 in normalized flux.} The second case study is the identification of the elemental source of missing spectral lines in the synthetic modelling. A mock dataset is used to show that absorption lines can be recovered when they are absent in one of the domains. This method can be applied to other fields, which use large data sets and are currently limited by modelling accuracy. The code used in this study is made publicly available on github\footnote{ \href{https://github.com/teaghan/Cycle_SN}{https://github.com/teaghan/Cycle\_SN}}.\\
\end{abstract}

\input{1_introduction}
\input{2_motivation}

\input{3_method}
\input{4_results}
\vspace{0.5cm}
\input{5_discussion}

\input{6_conclusion}

\acknowledgments

TO and SB acknowledge the support provided for a portion of this research by the Natural Sciences and Engineering Research Council of Canada (NSERC) Undergraduate Student Research Awards (USRA). YST is supported  by the NASA Hubble Fellowship grant HST-HF2-51425.001 awarded by the Space Telescope Science Institute. KV and SB acknowledge funding from the National Science and Engineering Research Council Discovery Grants program and the CREATE training program on New Technologies for Canadian Observatories.

\vspace{5mm}

\appendix
\input{7_appendix}

\clearpage

\bibliography{references}{}
\bibliographystyle{aasjournal}

\end{CJK*}
\end{document}

%% file: macros.tex
\newcommand\teff{T$_{\mathrm{eff}}$}
\newcommand\logg{log\,{\it g}}

\newcommand\eg{{\it e.g. }}
\newcommand\ie{{\it i.e. }}
\newcommand{\loss}{\mathcal{L}}
\newcommand{\xone}{$\mathcal{X}_{synth}$}
\newcommand{\xtwo}{$\mathcal{X}_{obs}$}

\newcommand{\Xonetwo}{$\mathcal{X}_{synth\rightarrow obs}$}
\newcommand{\Xtwoone}{$\mathcal{X}_{obs\rightarrow synth}$}

\newcommand{\z}{$\mathcal{Z}$}

\newcommand{\synth}{{synth}}
\newcommand{\obs}{{obs}}

\newcommand{\bxs}{\mathcal{X}_{\synth}}
\newcommand{\bxo}{\mathcal{X}_{\obs}}

\newcommand{\dist}{d}

\newcommand{\Es}{E_{\synth}}
\newcommand{\Ds}{D_{\synth}}
\newcommand{\Eo}{E_{\obs}}
\newcommand{\Do}{D_{\obs}}
\newcommand{\Esh}{E_{sh}}
\newcommand{\Dsh}{D_{sh}}
\newcommand{\Eshp}{E_{sh{/}p}}
\newcommand{\Dshp}{D_{sh{/}p}}
\newcommand{\Esp}{E_{sp}}
\newcommand{\Dsp}{D_{sp}}
\newcommand{\Cs}{C_{\synth}}
\newcommand{\Co}{C_{\obs}}

\newcommand{\Tso}{T_{\synth\rightarrow\obs}}
\newcommand{\Tos}{T_{\obs\rightarrow\synth}}
\newcommand{\Too}{T_{\obs\rightarrow\obs}}
\newcommand{\Tss}{T_{\synth\rightarrow\synth}}

\usepackage{color}
\definecolor{orange}{rgb}{1,0.5,0}
\definecolor{mydarkcyan}{rgb}{0,0.5,0.5}

\newcommand{\secref}[1]{Section~\ref{sec:#1}}
\newcommand{\Sec}[1]{Section~\ref{sec:#1}}
\newcommand{\Eq}[1]{Equation~\ref{eq:#1}}

\newcommand{\Fig}[1]{Figure~\ref{#1}}
\newcommand{\csn}[0]{\textsc{cycle-starnet }}
\newcommand{\csnn}[0]{\textsc{cycle-starnet}}
\newcommand{\kv}[1]{{#1}}

\newcommand{\YST}[1]{{#1}}

\newcommand{\obriain}[1]{{#1}}

\DeclareMathOperator*{\argmin}{arg\,min}

%% file: 1_introduction.tex
\section{Introduction}
\label{sec:intro}

Using theoretical models to decipher stellar spectra in terms of stellar properties is difficult. It requires detailed modeling of the photospheric surface layers, understanding a myriad of atomic and plasma processes, and calculating the radiative transfer through complex stellar atmospheres. Nevertheless, stellar spectra are one of the most important data sources that we have to understand stars.

Classically, methods that compare theoretical stellar spectra directly to observations have been used for decades. The comparison is typically performed manually \citep[{\it e.g.},][]{sneden2008,  aoki2013hires, venn2020}, but  massively multiplexed and higher resolution stellar spectroscopic surveys have been launched in the past few years \citep{gilmore2012gaiaeso, dalton2014, buder2018, holtzman2018}, where the data analysis approaches have started to become more automatic \citep{yanny2009segue, ting2017, ting2019, fabbro2018, zhang2019, bialek2020, guiglion2020}. 

Unfortunately, \obriain{methods that employ synthetic spectral models} typically suffer from the differences between theory and practice, referred to as the ``synthetic gap.'' The synthetic gap can be induced by both theoretical systematics and instrumental factors. In terms of theoretical systematics, many assumption are made during spectral modeling; {\it e.g.}, stellar atmospheres are often assumed to be one-dimensional, in hydrostatic equilibrium, and in local thermodynamic equilibrium. These assumptions often fail, causing systematic offsets between theoretical spectra and actual measurements. Instrumental factors can further introduce signatures that are not reproduced by theoretical modelling; {\it e.g.}, telluric lines imposed by the Earth's atmosphere and the image formation on the detector due to the light path through the telescope. Modern spectroscopic pipelines \citep[{\it e.g.},][]{ballester2000, opera2012} often need to make accurate assumptions about the instrumental signatures in order to reproduce realistic and consistent spectra.  These assumptions can contribute to the synthetic gap and limit the capabilities of the \obriain{methods that use synthetic spectra directly}.

Previous work has been done to attempt to overcome the synthetic gap between theory and observations. For instance, efforts have been made towards incorporating non-LTE and 3D hydrodynamic effects in the model atmospheres \citep[{\it e.g.},][]{amarsi2015, amarsi2016, kovalev2019}. Other methods have been proposed to isolate spectral regions -- only using spectral regions where the astrophysics and instrumental effects are better understood \citep[{\it e.g.},][]{jahandar2017, ting2019}.  \kv{On top of that, \citet{coelho2020} compared the use of synthetic versus empirical libraries and studied the impact of spectral coverage versus missing physics.} Furthermore, some have attempted to reduce this gap by augmenting the synthetic data through sampling and adding noise to make the spectra look more realistic~\citep[{\it e.g.},][]{bialek2020}. Despite these efforts, there is still a large amount of room for improvement.

In contrast, methods that use the empirical observed data directly for the spectral templates have been proposed \citep{ness2015cannon, ting2017, fabbro2018, leung2019,Xiang2019, Sharma2015, Prugniel2011}. These ``data-driven'' methods skip the direct use of synthetic spectra, but depend on \emph{a priori} knowledge of stellar parameters and elemental abundances for a large training set. Accordingly, these methods learn a model that directly translates spectra into physical characteristics, or vice versa. Due to the increasing number of amassed spectra, training these data-driven methods has become more tangible.

However, the stellar labels determined from data-driven models are limited in accuracy. \obriain{By training on previously derived stellar parameters and abundances, the errors in the original spectra and their parameters are propagated through to the final results. Therefore, the systematic errors are difficult to uncover when comparing the predictions to the original pipeline results. Naturally, this raises doubts on whether the model is learning actual physics or simply inheriting the biases from the original pipeline.} Lastly, building data-driven models requires high quality data for training, often in the form of high signal-to-noise spectra. In most cases, collecting a sufficient number of high quality empirical templates that span the full range of the stellar parameter and elemental abundance ranges can be difficult, if not impractical. 

In this study, we propose a novel solution, \csnn, that overcomes the synthetic gap without suffering from the shortcomings of data-driven methods. At its core, \csn is trained to learn to transform data from the synthetic domain to the observed spectral domain. To accomplish this, the network leverages recent advancements in machine learning methodologies; specifically, Domain Adaptation~\citep{liu2017, zhu2017unpaired}. Furthermore, the method can directly work with noisy training spectra, and implicitly denoise them. In essence, the domain adaptation is accomplished by forcing the two domains to share an abstract representation, which we use to exploit the connections found between the domains.

\obriain{\csn shows how unlabelled observed data can be used to calibrate synthetic spectra to become consistent with the observations, while also correcting for any improperly modelled spectral features. We refer to this as \em{auto-calibration}.} \obriain{In other words, our approach alleviates the need to know stellar labels before training (as is the case for data-driven methods), and at the same time, bridges the synthetic gap that plagues existing spectral analysis methods, which use synthetic spectra directly.} 

This paper is organized as follows: In \Sec{motivation}, we detail the critical insights of \csnn. In \Sec{method}, the technical details of \csn are described. In \Sec{results}, \csn is used in two case studies; in particular, correcting the systematics in theoretical models through domain adaptation, and identifying missing spectral features in the synthetic models. In \Sec{discussion}, the advantages of \csn compared to other spectral analysis techniques are discussed, as well as its limitations. We conclude the study in \Sec{conclusion}.

%% file: 2_motivation.tex
\section{Motivation and Overview of \csn}
\label{sec:motivation}

The main goal of \csn is to learn the connection between two sets of unlabelled spectra, and how to ``morph'' from one domain to another. In other words -- adopting terminology from the area of Domain Adaptation in Machine Learning -- \csn can transfer spectral models from the synthetic domain to the observed domain, and by doing so, it corrects for the systematic errors in the synthetic modelling.

Domain adaptation has a long history in the field of Machine Learning. Additionally, with the advent of the Generative Adversarial Network (GAN), GAN-based domain adaptation has seen numerous successes. For example, \cite{zhu2017unpaired} built a domain transfer model that was capable of translating photo images from day time settings to night time settings (and vice-versa), while keeping the content in the photos the same. In our work, we apply a similar method to translate between two spectral domains. As an analogy, one can think of the robust spectral features as the content in the photo images (which both domains share), whereas the day/night ``context'' are the systematics that we want to correct for. 

\obriain{\cite{liu2017} accomplished their domain adaptation through the combined use of variational auto-encoders and adversarial learning (\ie the use of GANs). In short, auto-encoders accomplish the task of representation learning by creating a bottleneck through the use of a neural network. This forms a compressed representation of the samples, which can make it easier to find correlations within the dataset. Adversarial learning, however, is typically accomplished through the use of a generator and discriminator; the generator aims to produce realistic samples, while discriminator acts as the teacher to help the generator achieve this goal. The method presented in \cite{liu2017} utilizes auto-encoders to transfer samples from one domain to the other, while the adversarial learning ensures that the transferred samples are accurate.}

An important aspect of these domain adaptation methods (including ours) is that the data is unpaired across the two domains. In other words, we are provided with data from each domain ({\it e.g.}, the synthetic and observed spectra), but we do not have samples from one domain that correspond to samples in the opposite domain. This is unlike the other proposed data-driven models \citep[{\it e.g.},][]{ness2015cannon, Ting2017b, Xiang2019} which assume that the corresponding stellar labels of the observed spectra are known beforehand, and then learn the label-spectra translation through supervised learning. Instead, for unsupervised domain adaptation, this mapping can be built with unpaired data. Not requiring paired samples is ideal for future uses in stellar spectroscopy because -- when obtaining newly observed spectra -- no prior assumptions or knowledge are needed regarding the stellar parameters and elemental abundances of the stars. 

\input{fig1}

We will elaborate on the technical details of \csn in \Sec{method}; here, we focus on the insights. 
\obriain{The critical insight is that, while the samples in the two domains are unlabelled, they should occupy the same regions of the stellar label space\footnote{In this study, we define stellar label to be the stellar parameters and elemental abundances. We do not investigate the impact of some astrophysical line broadening effects that may vary between elements, {\it e.g.}, Stark broadening, hyperfine corrections, and radiative damping parameters.} if they represent the same underlying objects. Therefore, a common representation can be constructed for these spectra. In \csnn, this common representation is formed in what we call the \emph{shared latent-space}.}

In order to achieve this goal, there are two mission-critical components of \csnn: (a) a method to extract the underlying hidden abstractions (or latent space) of each domain, essentially performing a dimensionality reduction of spectra, and (b) an algorithm to ensure that the abstractions for the two domains are the same. \obriain{Similar to \cite{liu2017}}, the former is done by learning to extract information with two separate auto-encoders for the individual domains. As for the latter, the individual abstractions are made to be the same by forcing the latent spaces of the two domains to be ``shared''. This implies that the latent embeddings from one domain are also valid in the other domain. In practice, this means that the auto-encoder networks are able to reproduce samples within the same domain, as well as map samples from one domain to the other. This concept is visualized in \obriain{Figure~\ref{fig1}.}

\obriain{On top of this framework, inspired by \cite{gonzalez2018}, we apply a twist that tightens our method: we leave room for a non-shared latent space. In more detail, the synthetic domain is only able to show variations that are a subset of what happens in real data. For example, the observed data might have instrumental variations that are not seen in the synthetic spectra. As a result, without leaving room for a non-shared latent space, it is impossible for the framework to model phenomena such as instrumental variations. To mitigate this issue, we provide the framework with the freedom to use additional latent variables, the \emph{split latent variables}, which are unique to the observed domain and not connected to the synthetic domain.}

%% file: fig1.tex
\begin{figure}
\centering
\includegraphics[width=0.8\columnwidth]{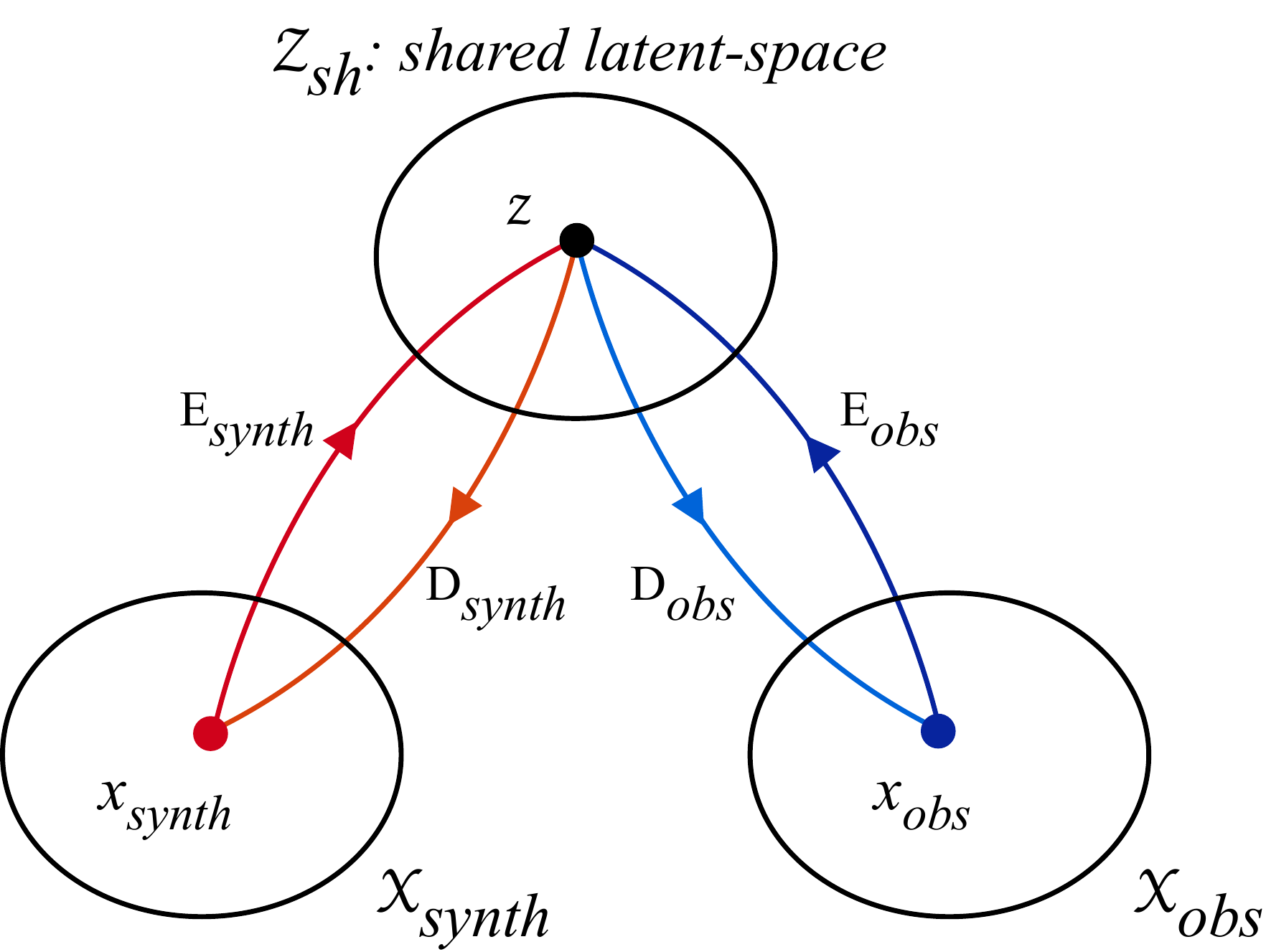}
\caption{A schematic diagram depicting the proposed method to provide a link between two domains of spectra: the synthetic domain, $\mathcal{X}_{synth}$, and the observed domain, $\mathcal{X}_{obs}$. This is accomplished by creating a shared space for the two domains to have a representation in, called the shared latent space, $\mathcal{Z}_{sh}$. The common space is created via autoencoders, where $E$ represents the encoder (or the dimensionality reduction component) and $D$ is the decoder (or the spectral generation component).
\label{fig1}}
\end{figure}

%% file: 3_method.tex
\section{Methodology}
\label{sec:method}

\obriain{In this section, we lay out the details of \csnn, which consists of two key components. The first component is a spectral emulator that produces synthetic spectra from a set of physical parameters. The second component is an unsupervised domain adaptation algorithm that transforms spectra from one spectral domain to another. We detail the two components in Sections \ref{sec:payne} and \ref{sec:method_cyclesn}, respectively.}

\input{fig2}

\subsection{Spectral Emulator}
\label{sec:payne}

\obriain{In order to provide a physically interpretable connection between the synthetic and observed spectra, the first step is to create a link between the physical parameters and synthetic spectra. In \csnn, we propose to accomplish this by including a synthetic emulator (see Figure~\ref{fig2}).  The synthetic emulator maps stellar labels, $\mathcal{Y}$, to the synthetic spectra, $\bxs$. Subsequently, the synthetic spectra can then be morphed into the observed domain via the domain adaptation network, which is explained next. The key here is to create a differentiable pipeline which can trace the latent representations back to the stellar labels.}

We adopt {\sc the payne} as our synthetic emulator \citep[for details, see][]{ting2019}, which utilizes a neural network as a physics surrogate to emulate the Kurucz ATLAS12/SYNTHE models \citep{kurucz1970, kurucz1981, kurucz1993, kurucz2005}. More explicitly, a multi-layer perceptron (MLP) network\footnote{In the original version in \citet{ting2019}, the authors adopted separate MLPs to emulate the flux variation of individual wavelength pixels independently. In this study, we adopt an improved version by considering a single large MLP to emulate the synthetic spectra as a whole. We found that using a single network facilitates the extraction of information from adjacent pixels, and thus improving the emulation precision. This version of {\sc the payne} can be found in the latest {\sc the payne} github: \href{https://github.com/tingyuansen/The_Payne}{https://github.com/tingyuansen/The\_Payne}} is trained on a set of Kurucz synthetic spectra, and the neural network learns how the spectral fluxes vary with respect to the stellar labels. \obriain{It is important to note that the synthetic emulator is a pre-trained network; the training of the domain adaptation does not involve the training of the emulator.}

\subsection{Domain adaptation}
\label{sec:method_cyclesn}

The primary motivation for \csn is the capacity to bridge the gap between observed and synthetic data sets using unsupervised methods; {\it i.e.}, learning hidden aspects of the two domains automatically, without any human supervision. For ease in explanation, we will refer to the domain of synthetic data as the ``synthetic domain'', and the observed data as the ``observed domain''. We further denote data from these two domains as $\mathcal{X}_{synth}$ and $\mathcal{X}_{obs}$, respectively. Note, however, that \csn is highly flexible, and can be used to transform between any two domains, and it is not restricted to transferring between synthetic and real observations. For example, one could also transfer between spectra obtained from different spectrographs, or between two different synthetic models, which we leave as future applications.

\subsubsection{Architecture}
\label{sec:architecture}

\obriain{In order to learn the mapping from one set of spectra to the other, we propose a method based on the UNsupervised Image-to-image Translation Networks \citep[\texttt{UNIT},][]{liu2017}. This framework is constructed out of encoders and decoders, as shown in Figure~\ref{fig2}.} One unique aspect of this architectural design is that we impose a hierarchical structure on the encoder-decoder pairs, which is used to facilitate training. Namely, two low-level encoder-decoder pairs ($\Es$-$\Ds$ and $\Eo$-$\Do$) are implemented that are dedicated to capturing domain-specific changes within spectra. We then utilize a high-level encoder-decoder pair ($\Esh$-$\Dsh$) that further abstracts the latent space; this pair is shared between both domains. Lastly, to implement the split latent space, we also make use of a second high-level encoder-decoder pair ($\Esp$-$\Dsp$) for the observed domain. \obriain{The shared and split latent spaces will be explained more thoroughly next.} Note that the dimensionality of the data gets reduced as it goes through the encoders -- thus the data is abstracted -- while the opposite happens when it goes through the decoders.

This architectural design allows low-level and data-related information to be learned by the domain-specific encoder-decoders. At the same time, the shared encoder-decoder learns to abstract high-level physical concepts that are shared amongst both domains. While this design choice was motivated by the architecture used for \texttt{UNIT} \citep{liu2017}, our framework is unique because of the use of the split encoder-decoder, which was found to provide improved convergence. Another key difference between \csn and \texttt{UNIT} is that we implement deterministic auto-encoders instead of variational auto-encoders in this study. The exact architectural design of these networks are outlined in \ref{sec:appendix_architecture}.

\subsubsection{The Latent Space}
\label{sec:shared_latent}

\paragraph{Shared latent space}

Using the framework outlined above, both datasets are encoded independently down to a shared representation. We denote this shared latent-space as $\mathcal{Z}_{sh}$, and illustrate this concept in Figure~\ref{fig1}. Once the shared space is created, a synthetic spectrum can be mapped to the latent-space, then the latent representation can be used to create the corresponding spectrum in the observed domain. Furthermore, as a bi-product of this shared latent-space, we create four domain mappings, which can be written as (1) $\mathcal{X}_{synth\rightarrow synth}$, (2) $\mathcal{X}_{obs\rightarrow obs}$, (3) $\mathcal{X}_{synth\rightarrow obs}$, and (4) $\mathcal{X}_{obs\rightarrow synth}$. Of these, the translation of $\mathcal{X}_{synth\rightarrow obs}$ (the mapping of imperfect synthetic models to the observed domain) is the main focus of this paper.

\paragraph{Split latent space}
\label{sec:split}

While the transfer of spectra between domains requires a shared latent space, not all of the information in one domain is present in the other. This is especially true when relating synthetic and observed spectra. For example, two stars in the observed domain may have the same set of stellar labels, yet vary in other characteristics. In other words, the two domains could have ``shared'' characteristics, but also unique properties of their own. The latent representation of the synthetic spectra can therefore be considered as a subset of the observed spectra, since the latter have other defining features from instrumental profiles ({\it e.g.}, line spread function) and observational affects ({\it e.g.}, telluric features) that might not be fully captured in the synthetic models. To account for this, in addition to the shared latent variables (that are common to both domains), we introduce a split latent-space, $\mathcal{Z}_{sp}$, which represents information that is unique to the observed domain. 

\obriain{In order to ensure that the shared and split information is contained in the appropriate latent-space, adversarial learning is applied to the latent-space, which is explained in \Sec{adversarial}. In short, if the unique observed information is encoded into the shared latent space, the adversarial training process will identify this and the encoded information will be adjusted.}

\subsubsection{Training the Network}
\label{sec:training}

In order for the network to be able to translate from one domain to the other, it should be able to perform -- simultaneously -- the following tasks:

\begin{enumerate}
	\item The encoder-decoders should be able to abstract and de-abstract spectra; meaning they should be able to map spectra to latent representations, then back to spectra within each domain.
	\item They should also be able to transfer spectra from one domain to the other. Once transferred, the spectra produced should look as if they are from the resulting domain.
	\item They should retain physical meaning within the transferred spectra. Therefore, when a spectrum is transferred from one domain to the other, it is once again transferred back to the original domain (thus forming a cycle). This cycled spectrum should be identical to the original spectrum.
\end{enumerate}

These three objectives are formed as three different loss functions, which we combine into a single loss function for optimization. We denote (1) the loss related to the within-domain reconstruction as $\loss_{rec}$, (2) the domain transfer loss as $\loss_{adv}$, and (3) the cycle-reconstruction loss as $\loss_{cr}$. Since these losses are replicated for both domains, the overall loss used during training can be written as
\begin{equation}
\begin{aligned}
    \loss = &\lambda_{synth}(\loss_{rec,synth} + \loss_{cr,synth}) \\ & +\lambda_{obs}(\loss_{rec,obs} + \loss_{cr,obs}) \\ &+ \lambda_{adv} (\loss_{adv,synth} +\loss_{adv,obs}),
    \label{eq:total_loss}
\end{aligned}
\end{equation}
where the $\lambda$s are hyper-parameters that control the influence of each term. 

This loss formulation is similar to that in \cite{liu2017}, except we adopt a mean squared distance, instead of a mean absolute distance. Each term is described below; for specific training details, see \ref{sec:appendix_training_details}.

\subsubsection{Within-Domain Reconstruction --- $\loss_{rec}$}
\label{sec:rec_task}

The first objective is to ensure that the encoder-decoders are able to reconstruct data within each domain. \obriain{If we denote the transfer of spectra from domain $a$ to domain $b$ to be $T_{a\rightarrow b}$, we can introduce the shorthand notations}
\begin{equation}
\begin{aligned}
    \mathcal{X}_{synth\rightarrow synth} 
    &\equiv \Tss\left(\bxs \right) \\
    &\equiv \Ds \Dsh \Esh \Es\left(\bxs \right),
    \\
    \mathcal{X}_{obs\rightarrow obs} 
    &\equiv \Too\left(\bxo \right)\\
    &\equiv \Do \Dshp \Eshp \Eo \left(\bxo \right).
\end{aligned}
\;.
\label{eq:reconstruction}
\end{equation}
Here, for simplicity, we have denoted using both encoder-decoders, $\Esh$-$\Dsh$ and $\Esp$-$\Dsp$, as $\Eshp$-$\Dshp$.
With this notation, the within-domain reconstruction loss function can be written as
\begin{equation}
    \loss_{rec} =\
    \dist\left(
    \bxs,\ 
    \mathcal{X}_{synth\rightarrow synth}
    \right)
    \\
    +
    \dist\left(
    \bxo,\  \mathcal{X}_{obs\rightarrow obs}
    \right),
\label{eq:ae_loss}
\end{equation}
where $\dist$ is the distance function. For the synthetic domain, a standard Mean Squared Error (MSE) loss is minimised, while for the observed domain, an MSE with samples weighted by the spectrum uncertainties (and bad pixels masked) is minimised.

\subsubsection{Cross-Domain Translation --- $\loss_{adv}$ }
\label{sec:adversarial}

\input{fig3}

As for the cross-domain translation, recall that there is no direct pairing between spectra in the two domains. To overcome this lack of direct pairing, as in \texttt{UNIT} \citep{liu2017}, we employ a generative adversarial network \citep[GAN,][]{goodfellow2014}. In other words, the translated\footnote{In this study, we use the words ``translated'' or ``transferred'' interchangeably. Both words designate the result of our domain adaptations, $\mathcal{X}_{synth \rightarrow obs}$ and $\mathcal{X}_{obs \rightarrow synth}$, which transforms spectra from one domain to the opposite domain.} spectra are required to ``look'' as though they belong in the translated domain.  Thus, we train critics (also referred to as discriminators) to distinguish between the actual spectra from each domain and the translated spectra.

A core idea behind GANs is that -- unlike typical deep network training where a fixed criteria exists -- the critic is also a deep network that is trained at the same time as the original generative network (here, our encoder-decoders). Furthermore, since the generative network is trained to fool -- whereas the critic is trained to discriminate -- the training process is an adversarial game; the critic minimizes a loss function, while the generative network maximizes it.

In the case of \csnn, as shown in Figure~\ref{fig3}, we use one critic for each domain: $\Cs$ and $\Co$. For both critics, they take in reconstructed\footnote{The critics are asked to discern the reconstructed spectra and the cross-domain spectra (instead of the original spectra and the cross-domain spectra) because this will facilitate the translation to denoise the cross-domain transferred spectra. Denoising will be discussed in \Sec{discussion}.} and cross-domain mapped spectra as inputs. The critics then predict a confidence value of whether or not each sample is ``real'' ({\it i.e.}, resembling the reconstructed spectra from the cross-domain) or ``fake'' (transferred from the opposite domain). Additionally, to help constrain the \obriain{shared} latent-representations to be the same for both synthetic and observed spectra, the critic networks act on the latent-space as well. More explicitly, the critics receive paired samples of spectra and their latent-representations, and predict a confidence value for each pair.

Mathematically, the training objective is defined with a binary cross-entropy function. Binary cross entropy assigns 0 and 1 values for the two groups (real versus fake). In our notation, we assign the value 1 for objects that truly belong to the group, and the value 0 otherwise. In short, the critic would want to assign value 1 for all objects that are reconstructed in a domain-specific manner, and 0 for objects that are passed through the cross-domain translation. If we denote the binary cross-entropy function as $H$, the loss for the critics can be summarized as

\begin{equation}
\begin{aligned}
    \loss&_{adv} 
    = 
    H\left(1, \Cs\left(\mathcal{X}_{synth\rightarrow synth}, \mathcal{Z}_{sh,synth}\right)\right)
    \\
    &+
    H\left(0, \Cs\left(\mathcal{X}_{obs\rightarrow synth}, \mathcal{Z}_{sh,obs}\right)\right)
    \\
    &+
    H\left(1, \Co\left(\mathcal{X}_{obs\rightarrow obs}, \mathcal{Z}_{sh,obs}, \mathcal{Z}_{sp,obs}\right)\right)
    \\
    &+
    H\left(0, \Co\left(\mathcal{X}_{synth\rightarrow obs}, \mathcal{Z}_{sh,synth},
    \mathcal{Z}_{sp,synth\rightarrow obs}\right)\right),
\end{aligned}
\label{eq:discrim_loss}
\end{equation}

\noindent
\obriain{where $\mathcal{Z}_{sp,synth\rightarrow obs}$ are the split latent representations produced when spectra are cycled from the synthetic domain, to the observed domain, and back to the synthetic domain.} Here, we have again used the short-hand notation for the transfer functions:
\begin{equation}
\begin{aligned}
    \mathcal{X}_{synth\rightarrow obs} 
    &\equiv \Tso\left(\bxs \right)\\ 
    &= \Do\Dshp\Esh\Es\left(\bxs \right),
    \\
    \mathcal{X}_{obs\rightarrow synth} 
    &\equiv \Tos\left(\bxo \right) \\ 
    &= \Ds\Dsh\Esh\Eo\left(\bxo \right).
\end{aligned}
    \;.
\label{eq:transfer}
\end{equation}

Note that the $\Dshp$ also requires the split latent variables, $\mathcal{Z}_{sp}$. Therefore, for each transferred spectrum from the synthetic to the observed domain, we choose a random observed spectrum, $\bxo$, encode it with $\Esp\Eo\left(\bxo \right)$, and use its split latent values for this process. Since we do not aim for a ``target'' observed spectrum -- but rather, to construct a realistic ``observed'' spectrum -- for the adversarial training, any $\bxo$ could be used and this would not harm the generality of $\loss_{adv}$.

While the task of the critics is to minimize this particular loss function, the task for the cross-domain auto-encoders is the complete opposite: to fool the critics. Therefore, the adversarial objective for the generative processes is to maximize $\loss_{adv}$, instead of minimizing.  In order to accomplish this, we simply switch the target class for the domain transferred spectra. The reconstructed ``true'' spectra are not used for this particular optimization process. Training of the critic network is alternated with the training of the rest of the framework, forming a min-max training setup as in typical GANs. For more details on the critic training, we refer readers to \cite{goodfellow2014}.

\subsubsection{Cycle-Reconstruction --- $\loss_{cr}$}
\label{sec:crec_task}

Accomplishing the within-domain reconstruction and cross-domain adversarial objectives would result in having a model that can -- not surprisingly -- reconstruct and cross-domain transfer spectra. In addition, the constraint of applying the critics on the latent-representations further provides the basis for a shared latent-space. However, there is no guarantee that for a given spectrum, $\mathcal{X}_{synth}$, the cross-domain generated spectrum, $\Tso(\mathcal{X}_{synth})$, is the correct corresponding spectrum in the \xtwo \ domain. Therefore, as described in \cite{liu2017}, we enforce that the physical meaning is preserved throughout the transfer by introducing a cycle-consistency constraint. In other words, we require that a spectrum from \xone \ can be mapped to the observed domain and then back to the synthetic domain accurately. The same applies to spectra in the \xtwo \ domain. 

Mathematically, we introduce the shorthand notations
\begin{equation}
\begin{aligned}
    \mathcal{X}_{synth \rightarrow obs \rightarrow synth} &= \Tos \circ \Tso\left(\bxs\right)
    \\
    \mathcal{X}_{obs \rightarrow synth \rightarrow obs} &= \Tso \circ \Tos\left(\bxo \right)
\end{aligned}
\;,
\label{eq:cycle}
\end{equation}

Importantly, when transferring an observed spectrum to the synthetic domain, the information in the split latent-space is lost. Therefore, to accurately cycle-reconstruct this spectrum, the originally encoded split latent variables are used when mapping this spectrum back to the observed domain.

We can write the cycle-reconstruction loss as
\begin{equation}
\begin{aligned}
    \loss_{cr} =& 
    \dist\left(
    \bxs,\
    \mathcal{X}_{synth \rightarrow obs \rightarrow synth}
    \right)
    \\
    &+
    \dist\left(
    \bxo, \mathcal{X}_{obs \rightarrow synth \rightarrow obs}
    \right)
\;,
\end{aligned}
\label{eq:cc_loss}
\end{equation}
which is similar to $\loss_{rec}$, but with \obriain{the cycled spectra}.

\input{fig4}

\input{fig5}

\input{fig6}

\subsubsection{Extracting Spectral Feature-Label Correlations}
\label{sec:correlation}

With the synthetic spectral emulator included, \csn provides a closed connection between stellar labels and the systematics-corrected translated spectra. This can be done by evaluating the continuous flow of $\mathcal{X}_{synth\rightarrow obs}=\Tso \circ Payne(\mathcal{Y})$, as shown in Figure~ \ref{fig4}.
Moreover, since each mapping -- including the synthetic emulator and auto-encoder -- is continuous, by interpreting the differential relations that the network has found, \textit{we can identify spectral features and associate them with individual stellar labels}.

Spectral features can be identified by calculating how individual input elemental abundances impact the output pixels in the observed domain. For example, if $\mathcal{X}_{synth\rightarrow obs}$ is the systematic-corrected model, then the partial derivative of the individual pixels with respect to a particular element, $\mathcal{Y}_j$, shows the spectral response to that element, which can be written as
\begin{equation}
\frac{\partial{\mathcal{X}_{synth \rightarrow obs}}}{\partial{\mathcal{Y}_j}} = \left(
\frac{\partial \mathcal{X}_{synth\rightarrow obs_{1}}}{\partial \mathcal{Y}_j},\ 
\cdots,\
\frac{\partial \mathcal{X}_{synth\rightarrow obs_{n}}}{\partial \mathcal{Y}_j} \right)
\end{equation}

The derivative spectrum for the synthetic domain emulator, $\mathcal{X}_{synth}=Payne(\mathcal{Y})$, can also be calculated, which provides the information held within our theoretical models. Therefore, by taking the difference between the two ``response functions'', the additional information that is not contained in the synthetic models can be revealed. This method is tested in \Sec{results_mocknewlines}.

%% file: fig2.tex
\begin{figure}
\centering
\includegraphics[width=\linewidth]{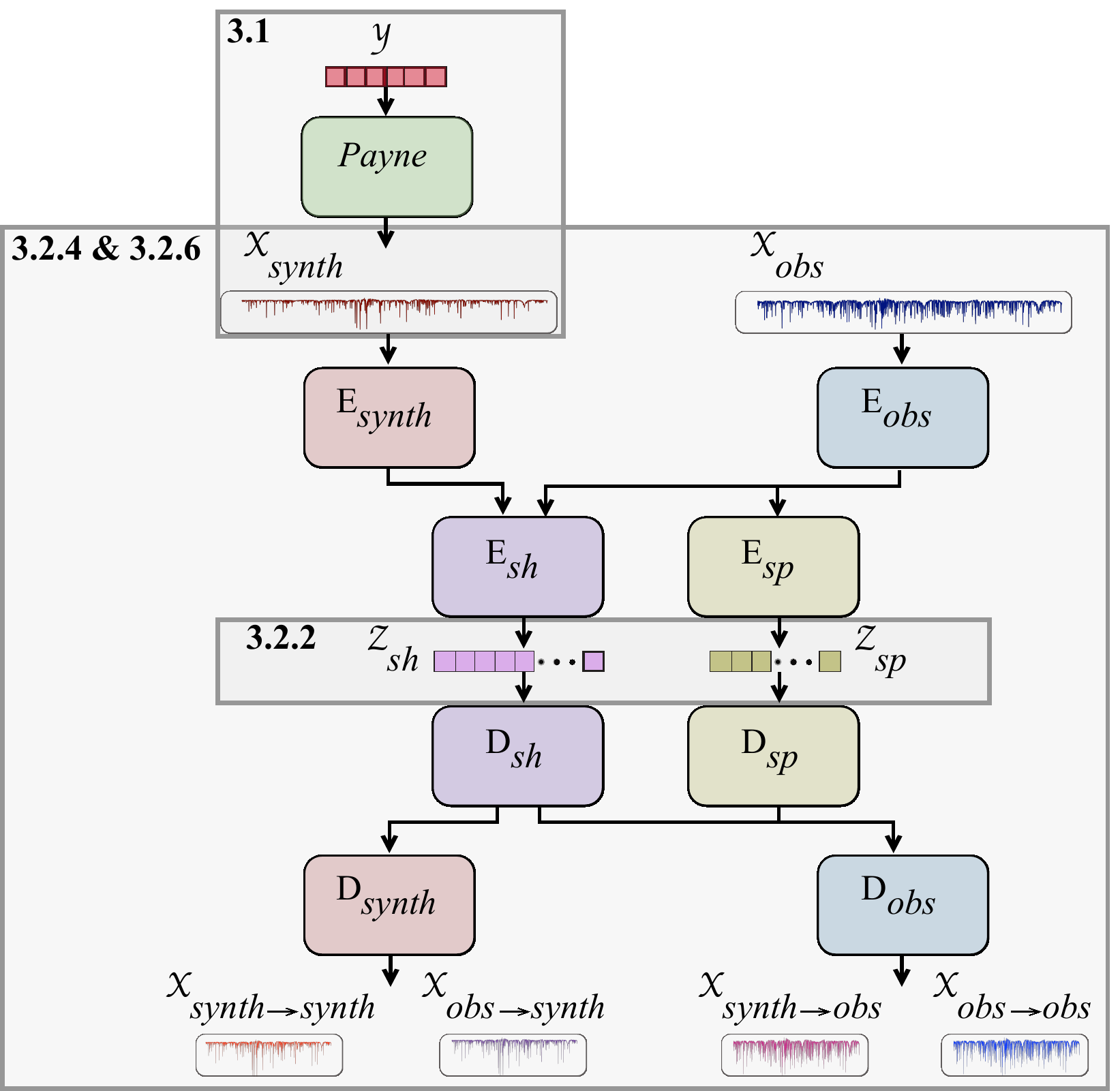}
\caption{
\obriain{A graphical representation of the \csn emulator and auto-encoder framework. First, a surrogate physics network ({\sc the payne}) emulates spectra, $\bxs$, from stellar labels, $\mathcal{Y}$. These synthetic spectra are then encoded through the encoder networks, $\Es$ and $\Esh$, to produce representations in the shared latent space, $\mathcal{Z}_{sh}$. Observed spectra are encoded in a similar manner (through $\Eo$ and $\Esh$), except an additional high-level encoder network, $\Esp$, is used to extract another representation in the split latent space, $\mathcal{Z}_{sp}$, which is unique to the observed domain. The shared latent representations can then be translated back to synthetic spectra through the use of the decoder networks, $\Dsh$ and $\Ds$. Alternatively, the shared latent representations can be combined with split latent representations to produce observed spectra by using the decoders, $\Dsh$, $\Dsp$, and $\Do$. These processes allow for spectra to be reconstructed within the same domain or transferred from one domain to another. Combining the emulator with the domain transfer network makes it possible to trace the latent representations back to the physical stellar labels. The individual components discussed in \Sec{method} are annotated in the plot.}}
\label{fig2}
\end{figure}

%% file: fig3.tex
\begin{figure}
\centering
\includegraphics[width=\linewidth]{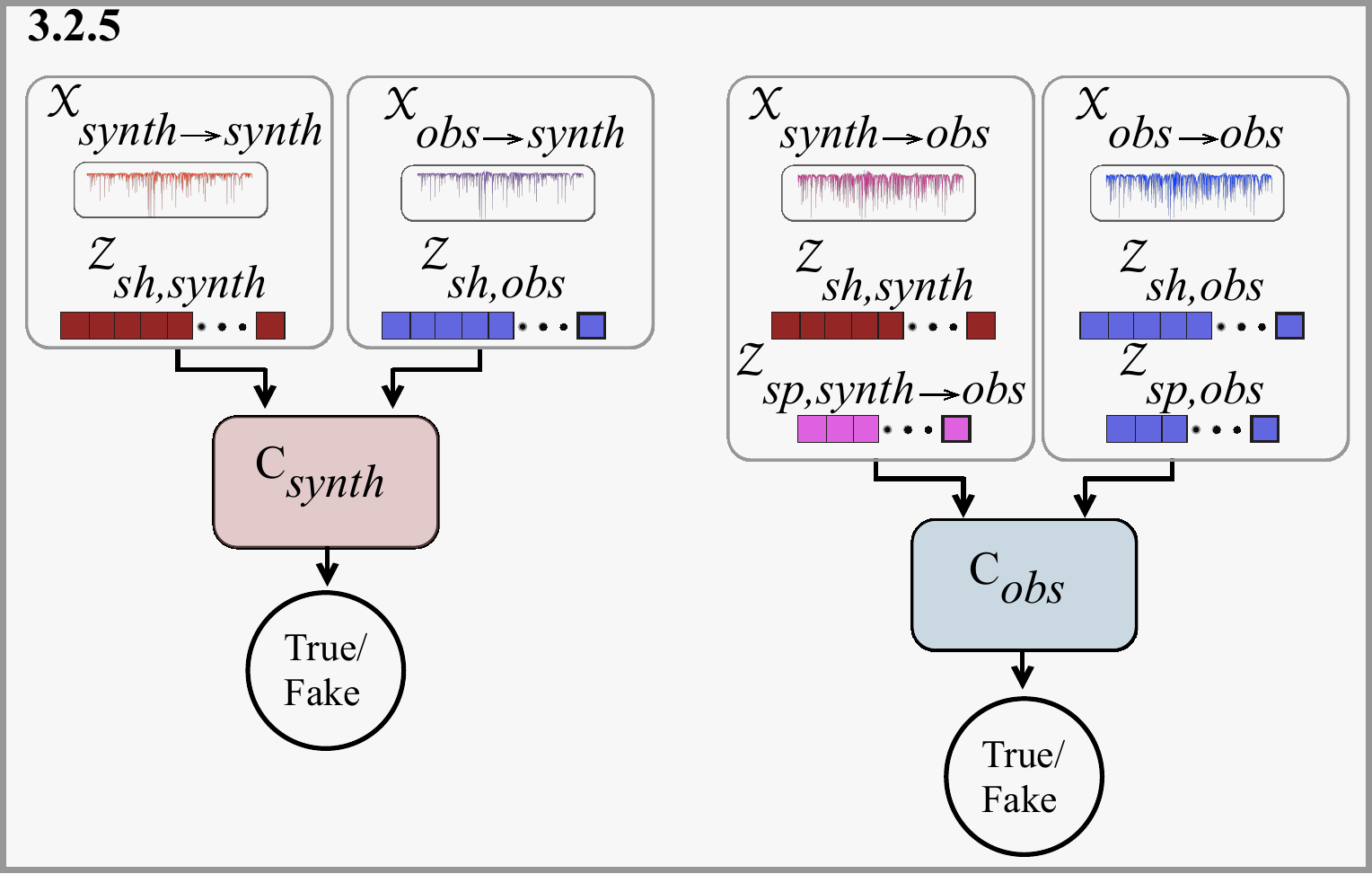}
\caption{
\obriain{A graphical representation of the \csn adversarial framework. One critic network is designated to each domain, $\Cs$ and $\Co$. Throughout training, these networks attempt to classify samples as ``true'' (the reconstructed samples) or ``fake'' (the cross-domain transferred samples) in order to provide information for the domain transfer process to make improvements. Combining the spectra with their associated latent representations -- as inputs to the critic networks -- helps to force both the domain transferred spectra to be accurate and the shared latent space to be truly shared between the two domains. This process is explained more thoroughly in \Sec{adversarial}.}}
\label{fig3}
\end{figure}

%% file: fig4.tex
\begin{figure}
\centering
\includegraphics[width=0.8\columnwidth]{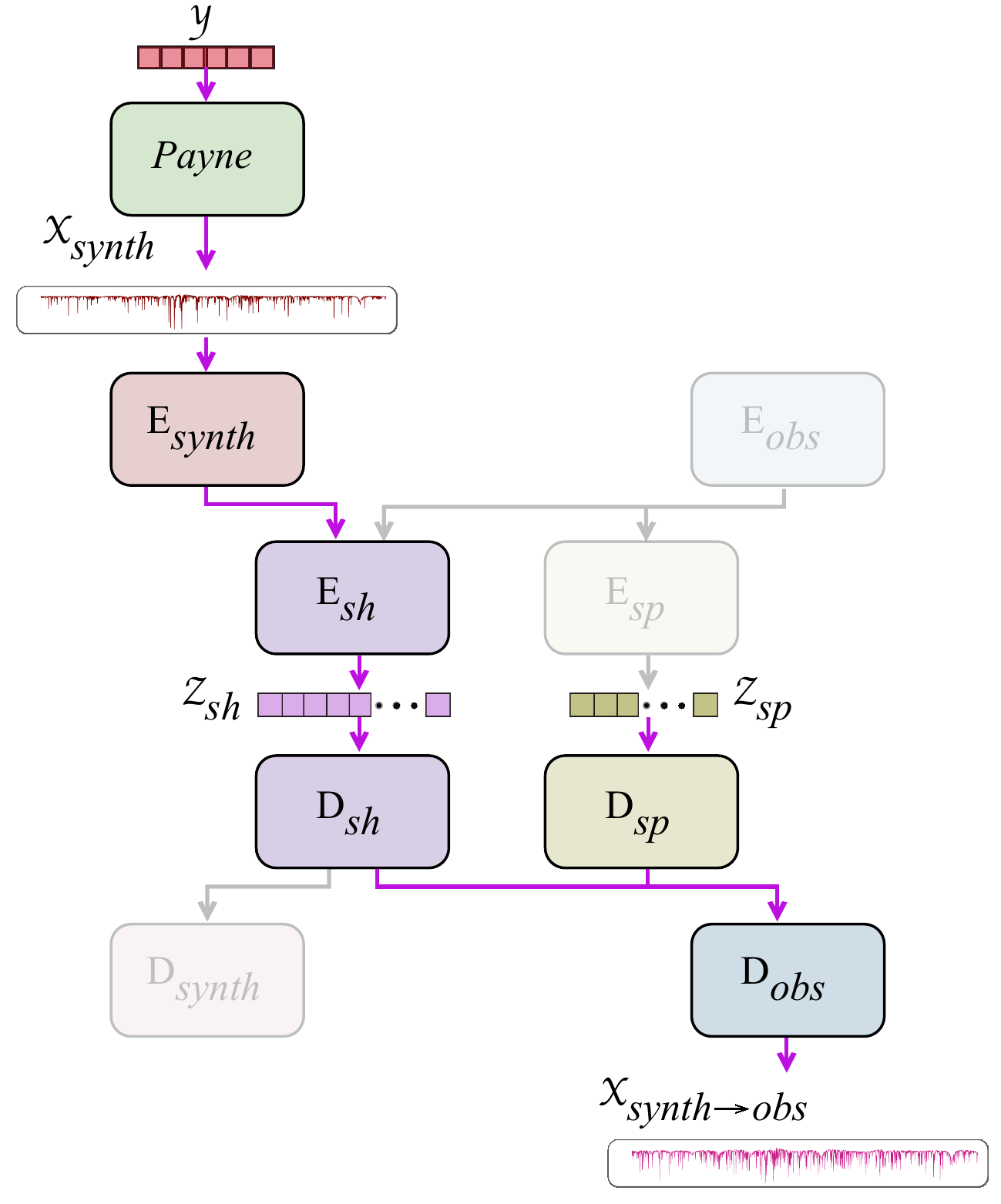}
\caption{A schematic diagram on generating systematic-corrected data-driven models. After \csn is trained, the network can create systematic-corrected spectra by mapping stellar labels, $\mathcal{Y}$, through the flow highlighted in this figure. Since the mapping is continuous, we can also associate spectral features to stellar labels by taking the differential derivatives of the network with respect to the input stellar labels.
\label{fig4}}
\end{figure}

%% file: fig5.tex
\begin{figure*}
\centering
\includegraphics[width=\linewidth]{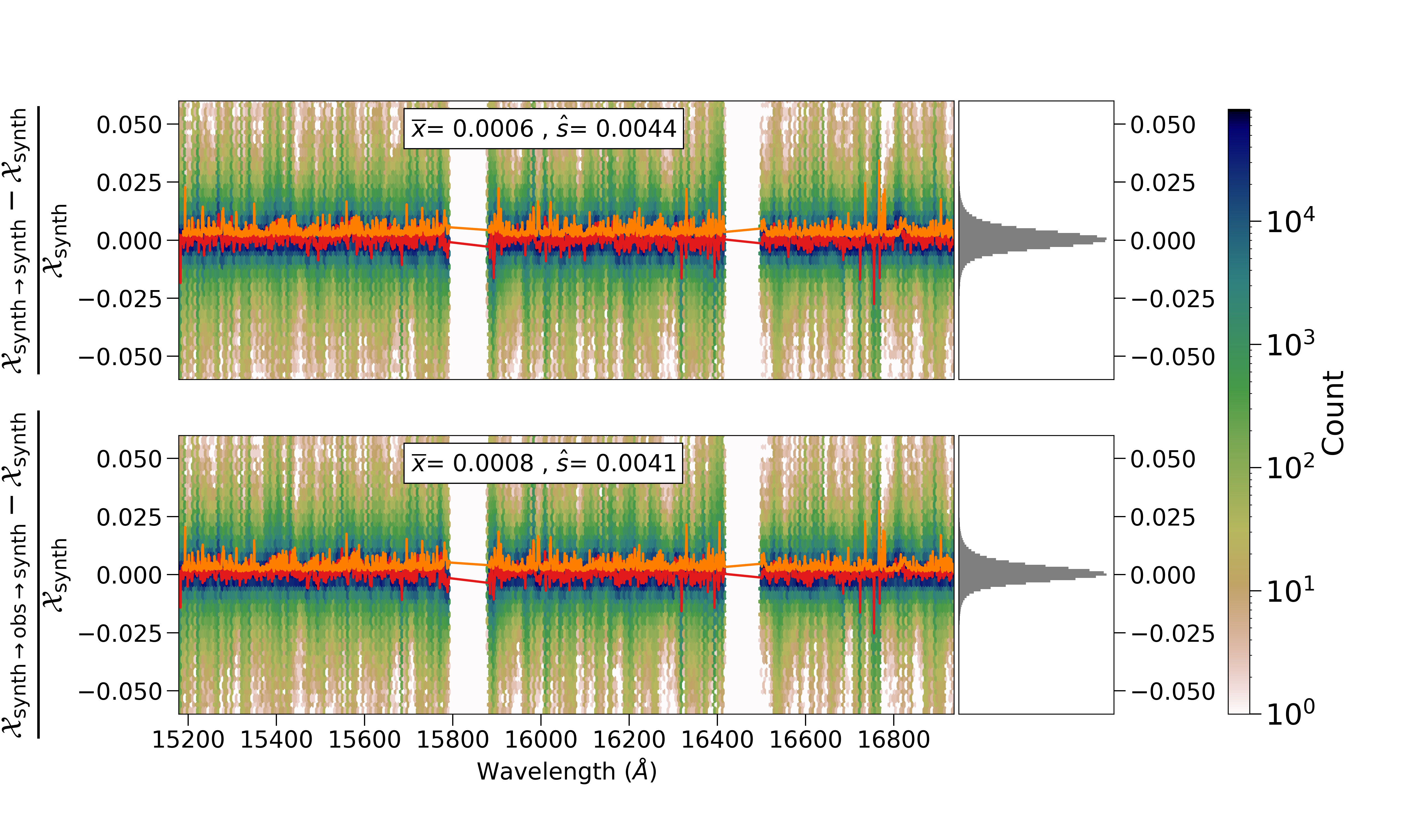}
\caption{The reconstruction of the synthetic domain. \csn transfers spectra through autoencoders, and the figure quantifies the \YST{reconstruction errors} from these autoencoders. Shown are the relative residuals for the test set of 7,000 synthetic spectra whose stellar labels are drawn randomly from the APOGEE-Payne catalog. \YST{Overplotted in each panel are the pixel-wise mean residual (red) and mean absolute residual (orange) with a $\sim$10\AA\ shift applied between the two to improve clarity.} The top panel shows the relative residuals between the reconstructed spectra $\mathcal{X}_{synth \rightarrow synth}$ and the original spectra \xone. Similarly, the bottom panel shows a comparison between the cycle-reconstructed spectra, $\mathcal{X}_{synth \rightarrow obs \rightarrow synth}$, and the original spectra, \xone. For each plot, the mean bias $\overline{x}$ and \YST{scatter} $\hat{s}$ are stated. \csn incurs a negligible \YST{reconstruction scatter} (0.4\%) and bias ($< 0.1\%$).
\label{fig5}}
\end{figure*}

%% file: fig6.tex
\begin{figure*}
\centering
\includegraphics[width=\linewidth]{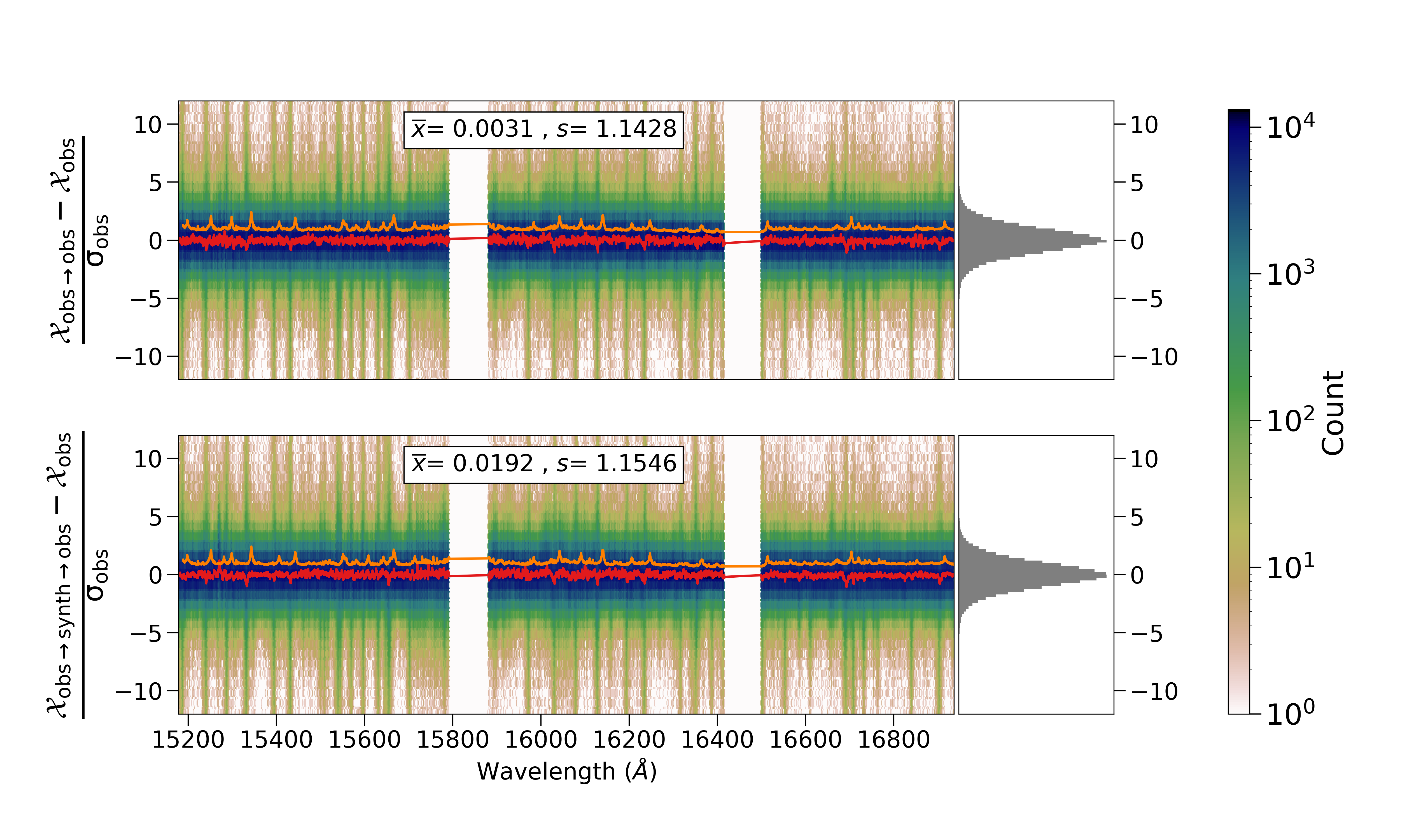}
\caption{Similar to Figure~\ref{fig5}, but here we show the reconstruction capabilities regarding the observed domain. However, since the observations are noisy, we do not expect the ``denoised'' reconstructions to be exactly the same as the input spectra. Therefore, we normalize the residuals with the reported uncertainties of APOGEE. \YST{Overplotted in each panel are the pixel-wise mean residual (red) and mean absolute residual (orange) with a $\sim$10\AA\ shift applied between the two to improve clarity.} The reconstructions have a \YST{normalized scatter} of $\sim 1$, demonstrating that the reconstruction in the observed domain is accurate and is consistent with the observational uncertainties.
\label{fig6}}
\end{figure*}

%% file: 4_results.tex
\section{Experiments \& Results}
\label{sec:results}

In this section, we present two case studies to show that \csn can generate systematic-corrected models from unlabelled observed spectra. In \Sec{results_domaintransfer}, we show how Kurucz synthetic models translate into APOGEE observed spectra. We quantify the agreement between spectra in terms of the residuals and via a t-SNE analysis. Admittedly, the accurate agreement of spectra does not necessarily guarantee that \csn has learned actual physics. Therefore, we further investigate the derivatives of \csn in \Sec{results_mocknewlines} to determine how the network derivatives have aligned with stellar labels. This second case study shows that the network has learned the actual physics behind spectra. Specifically, we show that \csn can associate missing spectral features in the synthetic spectra to their correct corresponding elemental abundances.

\subsection{Mitigating the synthetic gap with \csn}
\label{sec:results_domaintransfer}

In this section, we show how to mitigate the synthetic gap between the Kurucz models and APOGEE observations with \csnn.

\input{fig7}

\subsubsection{Experimental setup}

For this case study, {\sc atlas12}/{\sc synthe} models \citep{kur81,kur93,kur05,kur13} are adopted for the synthetic domain and {\sc the payne} is used as the synthetic emulator (\Sec{payne}). Similar to \citet{ting2019}, instead of using the default Kurucz line list, we utilized a calibrated line list by Cargile et al. (in prep.), which was tuned to better match the Solar and Arcturus FTS spectra. As for the observed domain, \YST{to be consistent with \citet{ting2019},} APOGEE DR14 spectra are adopted, which have been wavelength calibrated to vacuum to be consistent with the Kurucz models. These spectra are constructed by co-adding multiple velocity corrected visits of the same object. All spectra are continuum normalized using the same routine as in \citet{ting2019}.

For simplicity, the APOGEE-Payne catalog (the revised APOGEE catalog of stellar labels determined by using {\sc the payne}) is adopted as our reference. To reduce the effects of outlier spectra, only APOGEE spectra that have decent fits in the APOGEE-Payne catalog are included; \ie those with a reduced $\chi_R^2 < 50$ and a total broadening (which consists of both macroturbulence $v_{\rm macro}$ and rotation $v \sin i$) of less than 10 km/s. Finally, only APOGEE spectra with a median signal-to-noise of $30 < {\rm S/N}_{\rm pix} < 500$ are included to eliminate noisy and/or saturated spectra.

Furthermore, the APOGEE-Payne catalog is randomized to showcase how \csn can perform domain adaptation with unpaired spectra. In particular, for the observed domain, we adopt APOGEE spectra from half of the objects that meet the above criterion. For the other half of the objects, the APOGEE-Payne stellar labels are used to generate the synthetic Kurucz spectra. While the APOGEE-Payne labels are used to generate the Kurucz models and remove outlier spectra, we emphasize that \csn never ``sees'' the stellar labels of either domain; the training is entirely unsupervised. 

The 25 stellar labels from APOGEE-Payne, which are inherited include: $T_{\rm eff}$, $\log g$, microturbulence $v_{\rm turb}$, additional broadening $v_{\rm broad}$, 20 elemental abundances in [X/H], namely, C, N, O, Na, Mg, Al, Si, P, S, K, Ca, Ti, V, Cr, Mn, Fe, Co, Ni, Cu, Ge, and the isotopic ratio, C12/C13. The above procedure yields a set of 97,000 spectra in each domain. From these, we adopt 80,000 spectra as the training set, 10,000 spectra as the validation set, and withhold 7,000 spectra as the test set. Unless stated otherwise, all of the results shown below are based on the test set. 

\subsubsection{Within-domain reconstruction}

As discussed in Sections \ref{sec:rec_task} and \ref{sec:crec_task}, \csn provides two approaches for reconstructing spectra within the same domain: direct reconstruction and cycle-reconstruction. We first demonstrate that these within-domain reconstructions work, which are a necessary condition for the cross domain translations.

In Figure~\ref{fig5}, we show the accuracy of these mappings in the synthetic domain. The auto-encoded, $\mathcal{X}_{synth \rightarrow synth}$, and cycle-reconstructed spectra, $\mathcal{X}_{synth \rightarrow obs \rightarrow synth}$, are compared to the original spectra, $\mathcal{X}_{synth}$.  To show the relative residual, we normalize the difference with the original spectra, and display the 16-84 percentile as the 1$\sigma$ range \YST{to estimate the normalized scatter}. \YST{The red line demonstrates the pixel-by-pixel mean, and the orange line the absolute mean.} As demonstrated, \csn can reconstruct the synthetic domain with negligible bias ($<0.1\%$) and a scatter of  $\sim 0.4\%$. This applies to both the direct reconstruction and the cycle-reconstruction. Recall that the cycle-reconstruction is performed by passing information first to the opposite domain (here, the observed domain) and then back to the original domain (the synthetic domain). The fact that this cycle-reconstruction is accurate demonstrates that the latent space has learned not only information within-domain, but also information from the opposite domain.

Figure~\ref{fig6} shows similar results for the observed domain. However, since the observations are noisy, we do not expect the ``denoised'' reconstructions to be the exactly the same as the input spectra. Therefore, we normalize the residuals with the uncertainties of the observed spectra as reported by APOGEE, \YST{essentially calculating the reduced $\chi_R^2$}. The deviations for both reconstructions are minimal when compared to the original spectra, providing a negligible bias and a normalized scatter of $s \simeq 1$. This demonstrates that \csn is also able to reconstruct spectra in the noisy observed domain. Finally, the fact that the \YST{normalized scatter} is close to 1 demonstrates that the reconstructions are implicitly denoised. 

Nevertheless, outliers do exist and some pixels have more substantial deviations. There are $\sim 3\%$ of the pixels that have a normalized deviation greater than 3$\sigma$ ($s > 3$). The exact reason for these substantial variations is unclear, but we suspect some mischaracterizations of the APOGEE uncertainties may be a cause. Here, the uncertainties provided by APOGEE are assumed to be calibrated and uncorrelated between pixels, which might not be strictly true, especially for the resampled and co-added spectra. \YST{This also motivates us to take the 16-84 percentiles of the residuals, instead of the simple standard deviation, to calculate the reduced $\chi_R^2$ and normalized scatter.} 

\input{fig8}

\input{fig9}

\subsubsection{Domain adaptation with \csn}
\label{sec:results_domain_adapt}

\obriain{In order to evaluate the accuracy of the domain adaptation procedure, we perform a least-squares fitting on the observed spectra to determine how close the best-fit spectra match the observations. This provides a direct method to evaluate the improvements in model calibrations as well as estimates for the associated stellar labels (see \Sec{results_labels}). In more detail, as done in \cite{ting2019}, the fitting can be achieved by using the emulator directly, along with the error spectrum, $\sigma_{obs}$:}

\begin{equation}
\hat{\mathbf{y}} = \argmin_{\mathbf{y}} \left\|
    \frac{1}{\sigma_{obs}} \big( \mathcal{X}_{obs} - Payne(y)\big)
\right\|_2^2
\label{eq:paynefit}
\end{equation}

\obriain{With \csnn, there are two alternate ways to fit the spectra. The first is achieved by mapping the observed spectra to the synthetic domain and again use the emulator directly for the fitting:}
\begin{equation}
\hat{\mathbf{y}} = \argmin_{\mathbf{y}} \left\|
    \mathcal{X}_{obs \rightarrow synth} - Payne(y)
\right\|_2^2
\label{eq:synthfit}
\end{equation}

\obriain{Alternatively, the Kurucz models can be transferred to the observed domain; having their observed representations compared to the observations. In other words, \csn can be used to perform the least-squares fitting in the observed domain by forward modelling the synthetic spectra:}

\begin{equation}
\hat{\mathbf{y}} = \argmin_{\mathbf{y}} \left\|
    \frac{1}{\sigma_{obs}}\big(\mathcal{X}_{obs} - \Tso(Payne(y)) \big)
\right\|_2^2.
\label{eq:obsfit}
\end{equation}

\noindent
\obriain{Unless stated otherwise, we refer to \Eq{paynefit} as the Kurucz fitting and \Eq{obsfit} as the default \csn fitting. As we will discuss, for \csnn, it is better to forward model the synthetic spectra to the data space, instead of transforming the data to the synthetic space.}

In Figures~\ref{fig7}~and~\ref{fig8}, we show examples of how \csn is capable of adapting spectra from one domain to the other -- a key objective of this study. Figure~\ref{fig7} demonstrates this procedure for a typical M-giant with solar abundances in APOGEE. The upper panel compares the APOGEE spectrum to the corresponding best-fit Kurucz model. It is clear that, even though the two spectra are normalized with the same procedure, the synthetic gap persists. In contrast, the second panel of \Fig{fig7} compares the same APOGEE spectrum to the best-fit \csn model. The transferred model illustrates how \csn can correct for the improperly modelled spectral features.  Furthermore, \csn has learned to understand the imperfect continuum normalization in the data and produces transferred models that are consistent with such normalization. \obriain{The last two panels of \Fig{fig7} show the residuals for these comparisons across the entire wavelength region and normalized by the reported pixel uncertainties in the APOGEE spectrum..}

\YST{To further illustrate this point,} in Figure~\ref{fig8}, we analyze the residuals for all 7,000 test spectra. The top panel shows the residuals between the APOGEE spectra and best-fit Kurucz models, whereas the bottom panel shows these same residuals with the transferred synthetic spectra produced by \csnn. \obriain{The residuals are normalized by the reported APOGEE pixel-wise uncertainties in the observed spectra. Similar to Figure~\ref{fig7},} after the domain adaptation, the spectra exhibit much better agreement with the observations, reducing the synthetic gap. Evidently, when transferring the spectra with \csnn, the sample bias is 15 times smaller, and the sample \YST{reduced $\chi_R^2$} is 1.6 times smaller, reaching almost the same precision as the within-domain reconstructions, shown in Figure~\ref{fig6}. \obriain{Since the residuals are normalized by the uncertainties in the observed spectra, a normalized scatter (or reduced $\chi_R^2$) of $\sim 1$ demonstrates that the deviations are consistent with the observational uncertainties.} As previously mentioned, the Kurucz models used in this study have been generated with the improved line list by Cargile et al., (in prep.), as described in \citet{ting2019}. Consequently, the synthetic gap would be even larger if we were to use the original Kurucz models rather than those with the improved line list. 

\obriain{To determine how these residuals are dependent on the stellar parameters, Figure 9 shows a Kiel diagram coloured by the average reduced $\chi_R^2$ in the best-fit spectra. The left panel shows the results using the Kurucz models, and the right panel shows the results using the \csn calibrated models. While the Kurucz models exhibit larger deviations from the observations (particularly the cool M-giants), the best-fit spectra produced by \csn are in much better agreement and are consistent with the APOGEE observational uncertainties for all stellar types.}

\subsubsection{Visualization of domains via t-SNE}

An intuitive visualization of the synthetic gap -- and how we reduce it -- is provided by using t-Distributed Stochastic Neighbor Embeddings \citep[t-SNE,][]{maaten2008visualizing}. The t-SNE is a dimensionality reduction technique that is widely used to visualize high dimensional spaces.  Within the context of this paper, the t-SNE projects spectra from a $\sim$7000-dimensional ``spectral pixel'' space (or a 600-dimensional latent representation) to a compressed, 2-dimensional representation. This allows one to illustrate the distribution of a high-dimensional dataset in a 2D figure, where each sample is represented by a single point. Furthermore, the proximity of two points in the 2D space demonstrates the similarity of those two samples.  Since the t-SNE projected space is a lower-dimensional representation with arbitrary units, there are no explicit dimensions in the axes of the plots. 

In Figure~\ref{fig10}, using three separate t-SNE analyses, we show how our method mitigates the synthetic gap. The left panel shows the comparison between the Kurucz synthetic models, $\mathcal{X}_{synth}$, and the APOGEE spectra, $\mathcal{X}_{obs \rightarrow obs}$. Consistent with Figure~\ref{fig8}, the two domains are distinct in the t-SNE projection, which is further evidence of the synthetic gap. By contrast, the middle panel and the right panel illustrate the results of our domain adaptation. The middle panel compares the domain transferred spectra, $\mathcal{X}_{synth \rightarrow obs}$, to the auto-encoded version of the original observed spectra, $\mathcal{X}_{obs \rightarrow obs}$. Evidently, the domain adapted synthetic spectra show close agreement with the observed spectra, demonstrating that \csn can generate accurate synthetic spectra that are almost indistinguishable from those in the observed domain.

\input{fig10}

In these first two analyses, since \csn implicitly denoises spectra, we adopted the reconstructed versions of the observed spectra, $\mathcal{X}_{obs \rightarrow obs}$, as the reference for our comparisons instead of the original noisy spectra, $\mathcal{X}_{obs}$. The denoised versions were chosen to demonstrate that the synthetic gap is inherently due to imperfect modeling and reduction -- not the observational noise. Therefore, the auto-encoded $\mathcal{X}_{obs \rightarrow obs}$ is a more direct comparison with $\mathcal{X}_{synth}$ and $\mathcal{X}_{synth \rightarrow obs}$ because these are also noiseless.

Finally, the right panel shows the two latent representations of $\mathcal{X}_{synth}$ and $\mathcal{X}_{obs}$ produced by their respective encoders. The agreement between the latent representations demonstrates that \csn has indeed created a shared latent space that extracts common information from both domains.

\input{fig11}

\subsubsection{Deriving stellar parameters for APOGEE}
\label{sec:results_labels}

In this section, we study the recovery of stellar parameters from APOGEE spectra with the better-calibrated Kurucz models from \csnn. In particular, we fit 100,000 random APOGEE spectra taken from the observed domain and compare the results to the original stellar parameters inferred with {\sc the payne}, which uses the same Kurucz models. A more complete inference framework with detailed comparison to other pipelines -- as well as the study of elemental abundances -- is deferred to future work.

\obriain{As mentioned in \Sec{results_domain_adapt}, there are several methods that can be used to infer stellar labels from observed spectra. In \citet{ting2019}, the emulator is used directly as a surrogate for the physical model (\Eq{paynefit}). Naturally, this procedure is limited by the accuracy of the synthetic models. A second approach is to accommodate for these inaccuracies by producing synthetic counterparts for the observed spectra (using \csn) and, again, performing the fitting with the emulator (\Eq{synthfit}). Finally, the domain transfer -- achieved through \csn -- can be used to fit observed spectra directly in the observed domain (\Eq{obsfit}) by forward modeling the Kurucz models into the observed domain. In the following example, each of the objective functions (Equations \ref{eq:paynefit}, \ref{eq:synthfit}, and \ref{eq:obsfit}) are minimized using a second-order method (the Limited-Memory Broyden-Fletcher-Goldfarb-Shanno Algorithm, L-BFGS).}

\obriain{Figure~\ref{fig11} illustrates the differences between these three methods by plotting the respective estimates of surface gravity against effective temperature for the 100,000 APOGEE spectra. To provide a visual guide for the general expected trend, MIST isochrones at 7 Gyrs old are also plotted. In the top left panel, the results from \citet{ting2019} are shown. As discussed in their paper -- to omit poorly modelled spectral regions -- they constructed a spectroscopic mask, which is specific to the dataset. In the top right panel, the results produced by {\sc the payne} without using this spectroscopic mask are shown. Consequently, due to the imperfectness of the Kurucz models, fitting spectra without the spectroscopic mask exacerbates the systematic biases; for example, the cool M-giants.}

\obriain{The results from performing the \csn fitting in the synthetic space (using \Eq{synthfit}) are shown in the bottom left panel of Figure~\ref{fig11}. These estimates exemplify a tighter distribution of parameters (perhaps even overly tight compared to the isochrones), yet some metal poor dwarf stars are inconsistent with the expected trend.  While transferring observed spectra to the noiseless synthetic domain -- and performing a least-squares fitting on the synthetic spectra  -- corrects for several issues (\eg noise, outliers, and instrumental effects) it also limits the predicting power of the network. For instance, if an object is observed that is outside of the distribution of our synthetic training set, the observed spectrum may be moved to fit within this distribution when transferred to the synthetic domain. Alternatively, if we use the original observed spectrum, and instead transfer the synthetic spectra to the observed domain during the fitting procedure, we mitigate the potential for the underlying object to be altered. Furthermore, performing the fitting process in the observed domain can be cast as a likelihood estimation -- by taking into account the observed noise properly -- and is therefore more interpretable. Due to these considerations, we choose to adopt using \csn to transfer synthetic models to the observed domain as our preferred fitting routine throughout this paper.}

\obriain{The bottom right panel shows the results obtained by fitting with \csn in the observed domain; {\it i.e.} with the better-calibrated Kurucz models produced through domain adaptation (\Eq{obsfit}). Not surprisingly -- and consistent with Figure~\ref{fig8} -- the domain transferred models still provide a reduced amount of scatter in the estimated stellar parameters. Using \csn  to perform the fitting in the observed domain also shows improvements compared to fitting in the synthetic space. For instance, the distribution of \teff-\logg\ is more in line with the isochrones, and the problematic fits for the metal-poor dwarfs in the synthetic space are alleviated. Importantly, \csn allows the fitting to be performed on the entire spectrum, using all of the information collected, without the need for a spectroscopic mask.}

\subsection{Learning stellar physics with \csn}
\label{sec:results_mocknewlines}

While we demonstrate in Section~\ref{sec:results_domaintransfer} that \csn successfully morphs the Kurucz models to the APOGEE observations -- closing the synthetic gap -- it does not guarantee that \csn has extracted useful physics. In this section, we show a possible physical interpretation of \csn using the network's derivatives. In particular, the derivatives of individual flux intensities with respect to elemental abundances.

\subsubsection{Experimental setup}

As discussed in \Sec{correlation}, the derivatives (or ``flux responses'') of elemental abundances help determine which spectral features are associated with a particular element\footnote{This is not strictly true because some elemental abundances, especially those prolific electron donors, can also substantially change the stellar atmospheric structure, which indirectly affect all spectral features. As a result, spectral features that vary with a particular element do not necessarily imply that those features are due to the direct atomic/molecular transitions of that specific element.}. However, for the APOGEE observed spectra, it is impossible to know the ground truth; we simply do not know what may or may not be the missing in our line list. Therefore, as a proof of concept, we consider a mock ``observed'' data set drawn from the Kurucz models instead, which allows us to know the ground truth derivatives.  Applications to real spectra are deferred to future studies.

The training of this version of \csn is similar to the one in \Sec{results_domaintransfer}, however, this time we create a controlled observed domain. In more detail, instead of adopting the APOGEE spectra, an ``observed'' data set is synthesized with Kurucz models using the APOGEE-Payne labels that correspond to our observed training set used in \Sec{results_domaintransfer}. Noise is added to these mock observed spectra to mimic a more realistic observed training set.  

\obriain{Additionally, in the synthetic domain, we randomly masked approximately 30\% of the strong absorption features by setting them to the continuum level. In order to mask features that are associated with particular elemental sources, the gradients of the emulator network were investigated. In more detail -- for a solar metallicity K-giant ($T_{\rm eff}=4750\,$K, $\log g=2.5$) -- the gradient of each pixel in the simulated spectrum was calculated with respect to each input elemental abundance. Absorption lines with at least one pixel that had a gradient (in normalized flux) less than $-0.1\,$dex$^{-1}$ were considered ``strong features'' of that element, and surrounding pixels with a gradient less than $-0.02\,$dex$^{-1}$ were included as part of the feature. Furthermore, only elements with more than 3 strong features were included in the masking procedure.} 

To summarize, in this controlled experiment, two sets of unpaired Kurucz models are utilized. \obriain{The synthetic domain is composed of noiseless Kurucz models where 30\% of the strong spectral features (from prominent elements) are missing.} The observed spectra are the original Kurucz models without missing features, but with added noise, mimicking the real APOGEE spectra. This will demonstrate that \csn can learn actual physics. In particular, by learning from the data alone, \csn can correctly identify missing spectral features in the ``synthetic'' models and associate them to the correct corresponding elements.

The flux derivatives of the synthetic emulator, $\partial Payne(\mathcal{Y})/\partial \mathcal{Y}$, and the derivatives of the domain transferred spectra, $\partial \Tso(Payne(\mathcal{Y}))/\partial \mathcal{Y}$, are calculated. The former informs us of the original input line list in the ``synthetic'' domain. The latter reveals the ``true'', complete line list learned from the ``observed'' data. The differences between the two are used to identify the missing spectral features. 

\input{fig12}
\input{fig13}
\input{fig14}

\subsubsection{Recovering missing spectral features}

\obriain{An example of} the results of this domain adaptation problem are shown in Figure~\ref{fig12}. A K-giant ($T_{\rm eff}=4750\,$K, $\log g=2.5$) with solar abundances is used as our set of reference labels, $\mathcal{Y}$. The top panel shows the spectrum in the synthetic domain, $Payne(\mathcal{Y})$, and the second panel shows the transferred synthetic spectrum, $\Tso(Payne(\mathcal{Y}))$, as well as an actual observed spectrum with the same stellar labels. In this second panel, when mapped to the observed domain via \csnn, the missing features in the synthetic domain are correctly filled in. Similar to Figure~\ref{fig7}, this provides evidence that \csn is able to bridge the synthetic gap.

The last two panels in Figure~\ref{fig12} demonstrate that, not only can \csn accurately fill in the missing features, but it also associates them to their corresponding elements. In more detail, the third panel shows the difference between the transferred spectrum and the synthetic spectrum, which illustrates the missing lines and their corresponding elemental abundances. To make this more clear, we color-code the masked features with their associated elements; focusing on the four elements that have a prominent presence in the APOGEE H-band: Mg, Si, Fe, and C.  The final panel shows the differences between the true and recovered derivatives, as probed by \csnn.  In most cases, the differences in the derivatives are the strongest when calculated with respect to the correct input element.  This implies that the missing features are recovered with accurate associations, even though \csn was trained with noisy and unpaired observed spectra that mimic the APOGEE observations. Similar results are found for most other missing lines associated with other elemental abundances. Nonetheless, as occasionally seen  in Figure~\ref{fig12}, the gradients can be non-zero for unassociated elements, signaling that the unsupervised learning can still be improved in future studies. 

\obriain{In order to assess how well the network is able to recover missing lines for other objects, we performed a similar analysis on 100,000 different sets of stellar labels. Figure~\ref{fig13} exemplifies that the accurate recovery is consistent across the parameter space of the three atmospheric stellar parameters. In this figure, the varying areas of the parameter space are coloured by the percentage error in the pixel recovery for the missing lines. The percentage error is calculated by taking the absolute residual between the recovered masked pixels and ground truth flux (normalizing by the ground truth). The only regions which see a substantial discrepancy ($\sim$ 1\%) between the recovered lines and the ground truth are the coolest M-giants. This is presumed to be due to a lower number of training samples in this region.}

While this experiment produces encouraging results for line identifications, it also highlights some limitations. Specifically, \csn can struggle to identify the exact sources of features when more than one element contribute to the feature -- either directly or indirectly. This is especially true for the CNO molecular features \citep[via the molecular balance of CNO, {\it e.g.},][]{ting2018}. For instance, Figure~\ref{fig14} shows a few of the carbon and nitrogen features that were masked in the synthetic domain. As illustrated, \csn can assign the C and N related features to either C or N or both. In these cases, \csn will only be able to limit the potential sources rather than identify them precisely through a gradient analysis. However, such molecular features are usually very prominent in the spectrum, with many neighboring transitions within the wavelength region. As a result, incorporating prior domain knowledge of stellar spectroscopy could readily resolve this limitation. 

\input{tab1}

\obriain{Lastly, for the sake of completion, an experiment was conducted to analyze the extent of missing information in the synthetic domain that can be recovered. In more detail, nine instances of training \csn were completed while the percentage of masked absorption features was increased from 10\% to 90\% (in increments of 10\%), and the percentage error in pixel recovery for these masked regions was calculated. Table~\ref{table:masked_lines} summarizes these results by comparing the percentage error for the pixel recovery for each training instance. Interestingly, in terms of flux recovery, the network only appears to see a significant decrease in performance once there are 80\% of the strong features masked.} 

%% file: fig7.tex
\begin{figure*}
\centering
\includegraphics[width=0.9\linewidth]{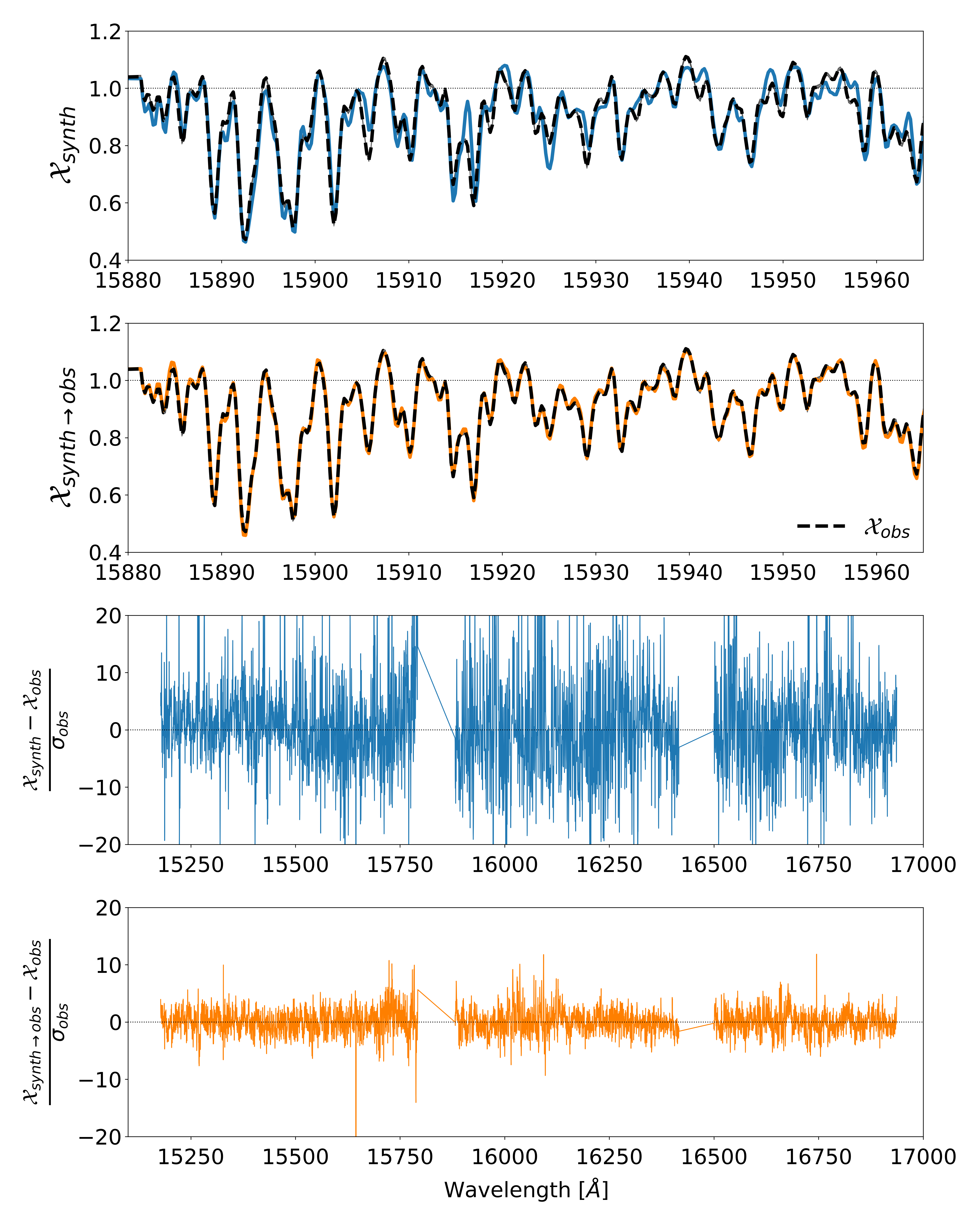}
\caption{Mitigating the synthetic gap with \csnn. The top panel shows a portion of the best-fitted Kurucz model for an M-giant APOGEE spectrum with solar-like abundances. While the Kurucz model is broadly consistent with the APOGEE observation, a minor synthetic gap persists. The second panel shows the \csn generated spectrum via domain transfer. Evidently, the transferred model exhibits better consistency with the observation, especially for cool stars like M-giants whose spectral models were not well calibrated. On top of that, \csn also learns the imperfect continuum normalization in the data (as illustrated by the fact that some normalized values are $> 1$) and produces transferred synthetic models that are normalized in a self-consistent manner. The last two panels show the normalized residuals for these same two comparisons, but across the entire wavelength range.
\label{fig7}}
\end{figure*}

%% file: fig8.tex
\begin{figure*}
\centering
\includegraphics[width=\linewidth]{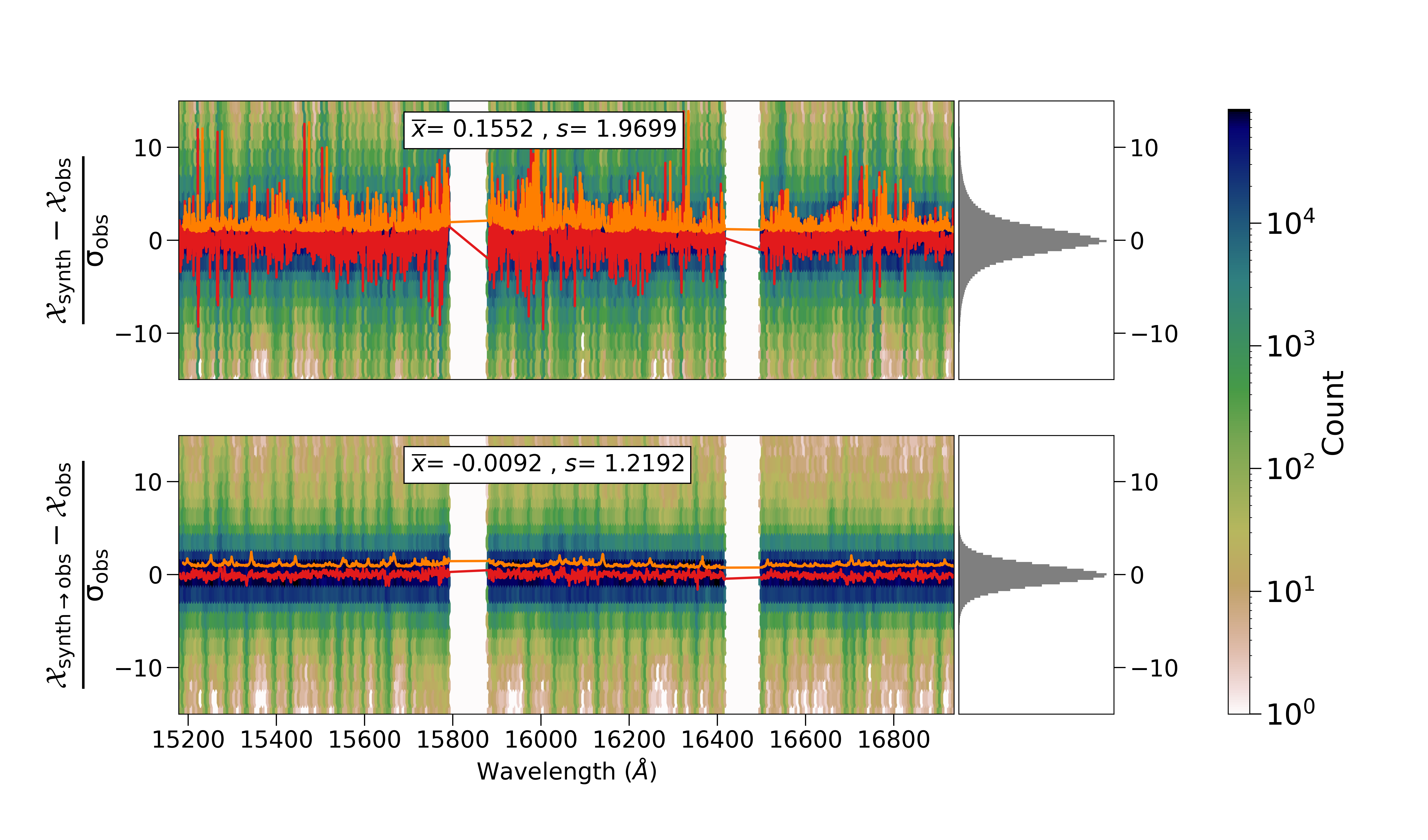}
\caption{\csn mitigates the synthetic gap between the Kurucz models and the APOGEE observations. The top panel shows the difference between the 7,000 APOGEE test spectra and their corresponding best-fit Kurucz models. Even with consistent continuum normalization, the residuals are more significant than the measured uncertainties with a non-negligible bias in the residuals; evidence of the synthetic gap. In contrast, the bottom panel shows a similar comparison between the APOGEE spectra and the Kurucz models that are transferred with \csnn. \YST{Overplotted in each panel are the pixel-wise mean residual (red) and mean absolute residual (orange) with a $\sim$10\AA\ shift applied between the two to improve clarity.} The transferred spectra demonstrate better consistency with the APOGEE observations and the residuals are largely consistent with the APOGEE reported uncertainties with a negligible bias.
\label{fig8}}
\end{figure*}

%% file: fig9.tex
\begin{figure*}
\centering
\includegraphics[width=\linewidth]{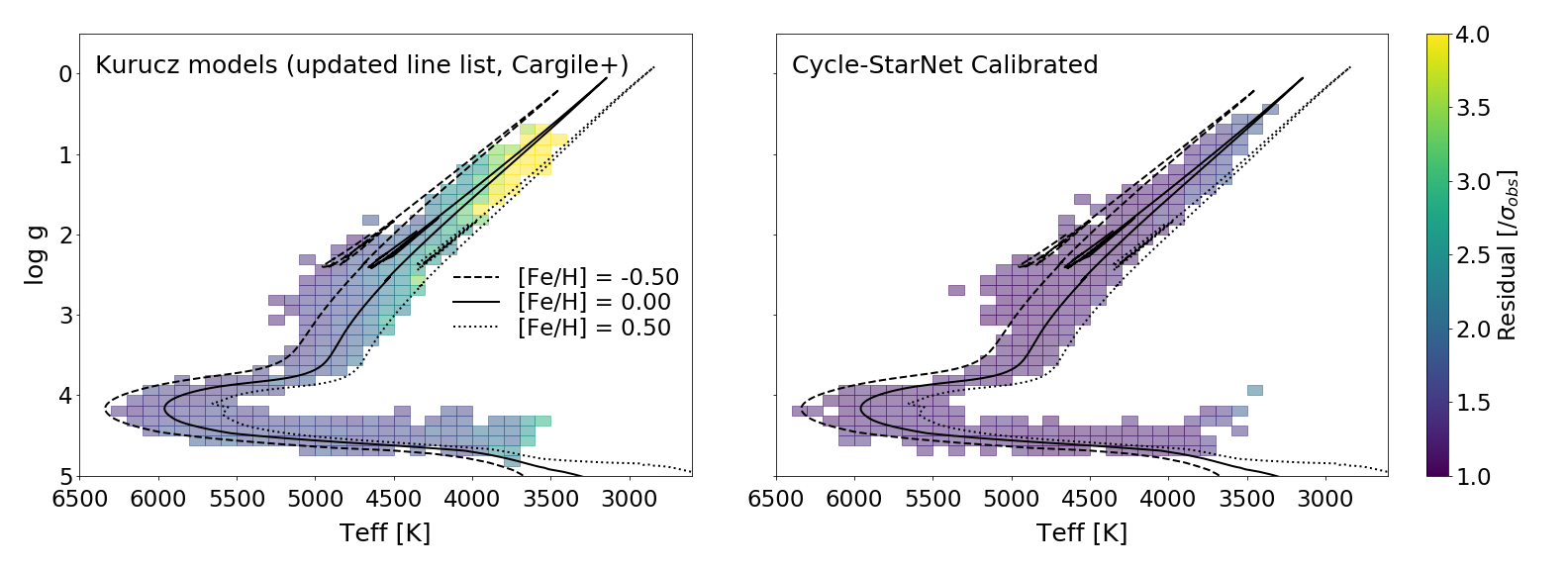}
\caption{\obriain{The dependence of the fitting residuals on the stellar parameters. The left panel shows the stellar labels derived using the original Kurucz models and is coloured by the average reduced $\chi_R^2$ (excluding outlying pixels) between the best-fit spectra and the observed spectra. We only show $T_{\rm eff}$-$\log g$ bins with more than 5 test spectra. The right panel shows the same comparison, but for stellar labels obtained using \csnn. MIST isochrones at 7 Gyrs old are also plotted to provide a visual guide.}
\label{fig9}}
\end{figure*}

%% file: fig10.tex
\begin{figure*}
\centering
\includegraphics[width=1.0\linewidth]{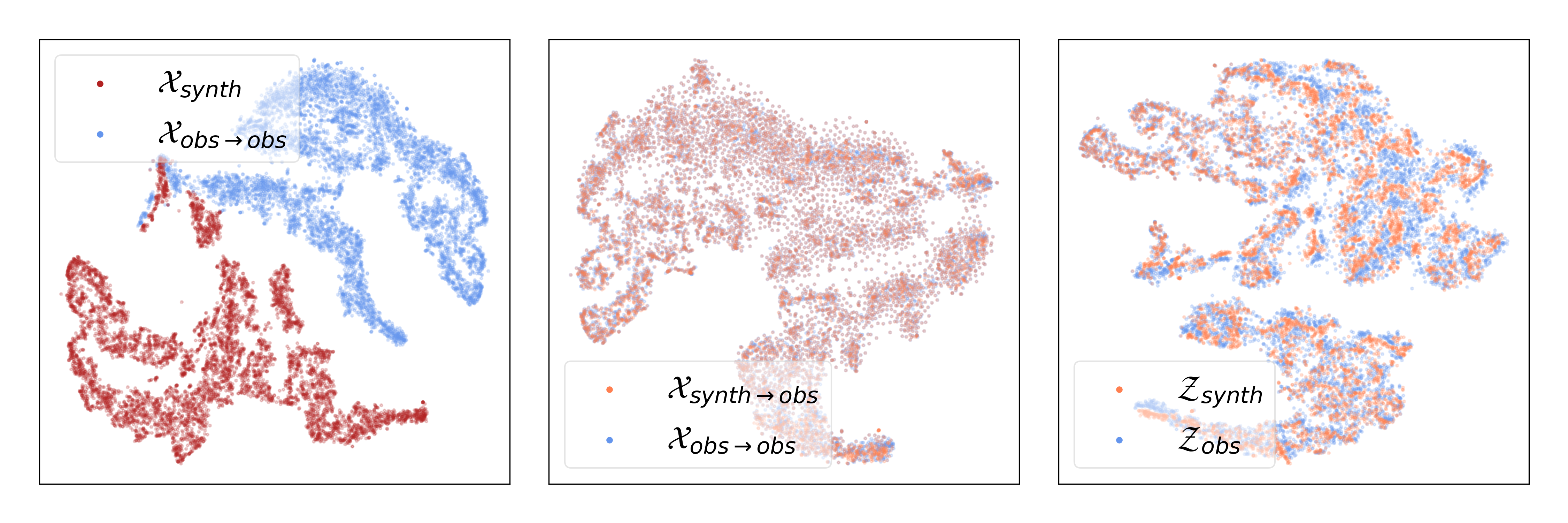}
\caption{The t-SNE projections of 7,000 pairs of test spectra. The left panel illustrates the synthetic gap between the Kurucz models and the APOGEE spectra; without domain adaptation, spectra from the two domains span different regions in the t-SNE projection. The middle panel shows the \csn results, which exemplifies the effectiveness of the generative aspect of \csnn. After domain adaptation, \csn successfully morphs the synthetic domain to the observed domain.  In both panels, to have a more robust comparison (see text for details), we consider the auto-encoded ``denoised'' version of the observed spectra, $\mathcal{X}_{obs \rightarrow obs}$, as our reference instead of the original version, $\mathcal{X}_{obs}$. The right panel shows the comparison of the latent representations, demonstrating that \csn created a common representation for the two domains. 
\label{fig10}}
\end{figure*}

%% file: fig11.tex
\begin{figure*}
\centering 
\includegraphics[width=1.0\textwidth]{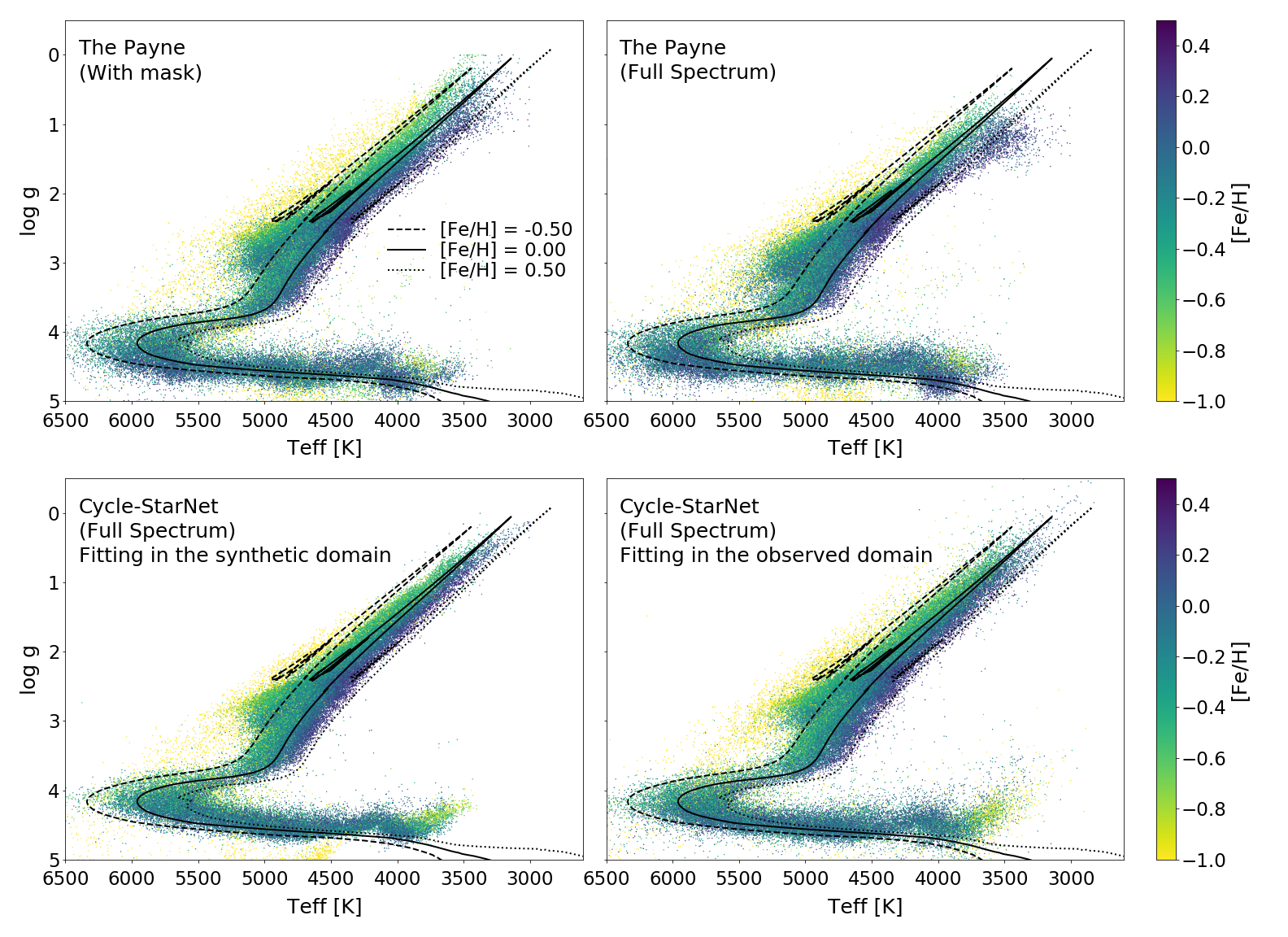}
\caption{Stellar parameters recovered by Cycle-StarNet for APOGEE spectra. Typical full spectral fitting techniques such as {\sc the payne} require extensive evaluation of the synthetic models to isolate wavelength regions where the models do not agree with the observation (top left panel). Using all pixels (top right panel) can exacerbate model systematics biases. \YST{The bottom panels show two ways of fitting spectra with \csnn.} \obriain{Either we transfer the observed spectra to the synthetic domain (\Eq{synthfit}), as shown in bottom left panel, or we directly translate the Kurucz models to the observed domain with Cycle-StarNet (\Eq{obsfit}), as shown the bottom right panel.}
\label{fig11}}
\end{figure*}

%% file: fig12.tex
\begin{figure*}
\centering 
\includegraphics[width=0.95\textwidth]{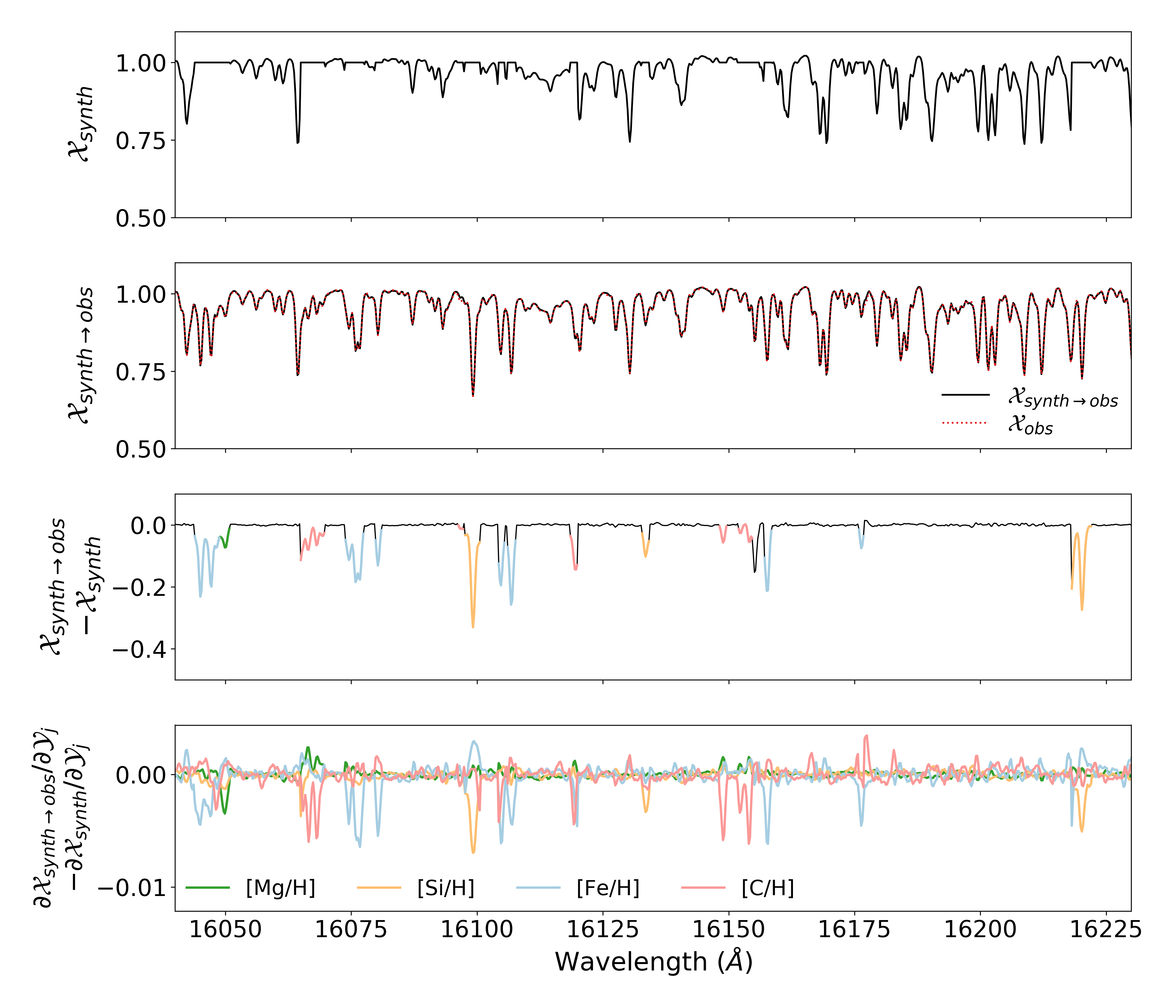}
\caption{ Identification of missing spectral features with \csnn. We present a mock study where we have perfect knowledge of both the synthetic and ``observed'' domains. We generate Kurucz spectral models, but mask 30\% of the spectral features in the ``synthetic'' domain. We then use the original Kurucz models (with noise added) as the observed domain. The top panel shows the Kurucz model for a Solar abundance K-giant with missing spectral features. The second panel compares the systematic-corrected transferred spectrum to the actual ``observed'' spectrum. The third panel shows the differences between the synthetic and transferred spectra, demonstrating the missing features; the missing features of Mg, Si, Fe, and C are annotated in green, orange, blue, and red, respectively. The final panel shows the differences in the \csn derivatives between the synthetic domain and the transferred models. The difference between the two demonstrates the additional information that \csn has learned from the observed domain, yet was not contained in the synthetic models. Even with noisy and unpaired observed spectra that mimic the APOGEE observations, \csn not only correctly recovers the missing features (the second panel), but it identifies the actual elemental sources of the missing spectral features, as shown in the final panel.
\label{fig12}}
\end{figure*}

%% file: fig13.tex
\begin{figure}
\includegraphics[width=\columnwidth]{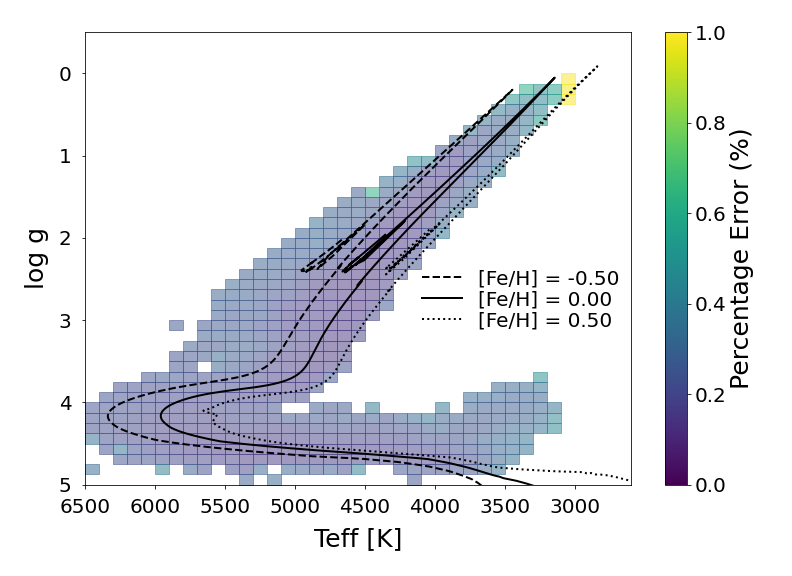}
\caption{\obriain{Missing line recovery as a function of \teff\ and \logg. Spectra in the synthetic domain (where 30\% of the strong lines are masked) are mapped to the observed domain. The recovery of lines are quantified by calculating the average percentage error between the recovered lines and the ``ground truth''. Except for the coolest M-giants (where the training samples are scarce), \csn can recover missing features across all stellar types.}
\label{fig13}}
\end{figure}

%% file: fig14.tex
\begin{figure*}
\includegraphics[width=0.95\linewidth]{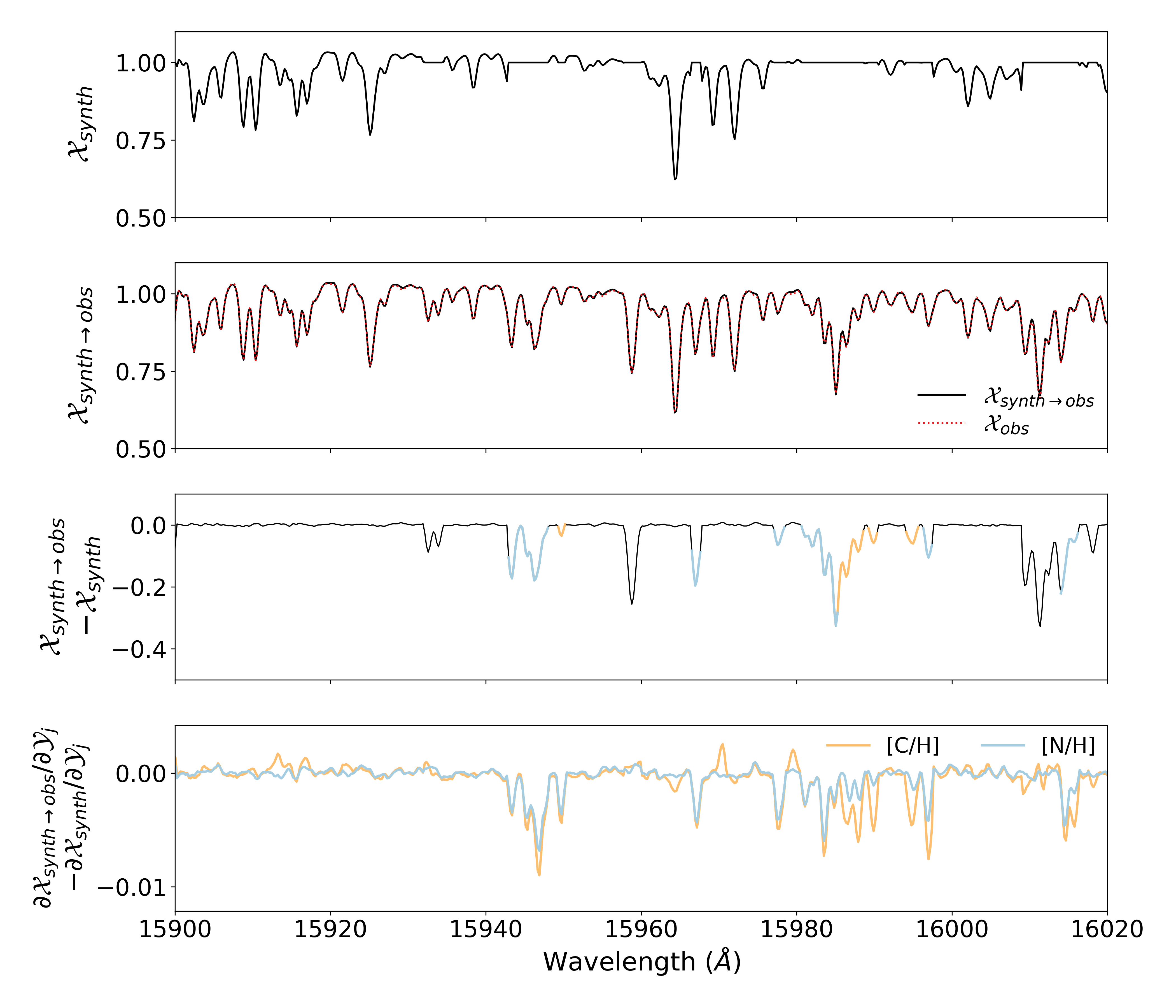}
\caption{Similar to Figure~\ref{fig12}, but for the carbon and nitrogen spectral features. While \csn still successfully identified missing features -- especially for the C and/or N related molecular features -- \csn sometimes inaccurately associates the features for being purely carbons or nitrogen or both. C and N are highly degenerate and can both contribute to the same features, directly or indirectly. In this case, \csn might struggle to distinguish the exact sources. Domain knowledge is needed to disentangle the exact sources.
\label{fig14}}
\end{figure*}

%% file: tab1.tex
\begin{table}[!htbp]
\caption{The average percentage error of pixel recovery for the masked absorption lines when varying the amount of masked features in the synthetic domain.
\label{table:masked_lines}}
\begin{tabular}{|cc|cc|}
\hline
\% of Features Masked & Avg. Error (\%) \\ \hline
10                 & 0.3835          \\
20                 & 0.4136          \\
30                 & 0.3547          \\
40                 & 0.3750          \\
50                 & 0.3862          \\
60                 & 0.3903          \\
70                 & 0.3794          \\
80                 & 4.4961          \\
90                 & 4.5906          \\ \hline
\end{tabular}
\end{table}

%% file: 5_discussion.tex
\section{Discussion}
\label{sec:discussion}

In this study, we showed how unsupervised domain adaptation is a promising method to auto-calibrate for model inaccuracies through exploiting large data sets. To illustrate this, we presented a case study using stellar spectroscopy surveys. In particular, we demonstrated that domain adaptation is a powerful idea that can harness the strengths of \obriain{large datasets} and \obriain{synthetic} modeling while mitigating their limitations. In the following, we first discuss the advantages of \csn when compared with standard spectral calibration and fitting methods. We then discuss the limitations of \csnn.

\subsection{Advantages of applying \csn}
\label{sec:hybrid}

A key advantage of \csn compared to existing data-driven models ({\it e.g.,} \citealt{ness2015cannon, fabbro2018, leung2019}), is that the training does not require any labels for the observed data. This is important as supervised training often inherits biases in the training labels. These biases are a result of obtaining training labels derived from other pipelines, which adopt their own set of models \YST{and fitting techniques} that are susceptible to systematics. In contrast, \csn learns a common abstraction of both the synthetic and observed domain -- purely from the data -- and directly adapts data in the synthetic domain to the observed domain and vice versa. Therefore, unlike other approaches, \csn is not subject to biases in the training labels. Thus, \csn provides an entirely new idea to construct data-driven models through unlabelled large datasets. Furthermore, one can interpret the shared representation as a multi-dimensional match, alleviating the need for spatial cross-matching between surveys. This would ultimately mitigate selection biases seen in pure data-driven analyses.

Moreover, \csn can yield robust and denoised data-driven models, even when trained with noisy spectra. We attribute this ``effective denoising'' of observed spectra to the following subtleties related to our method. First, the reconstruction and the cycle-reconstruction loss functions are weighted by the uncertainties in the spectra. As a result, this weighting forces the network to focus more on the cleaner portions of the dataset and down-weight the noisier spectral fluxes. In addition, when provided with enough data, auto-encoders learn to reconstruct the common information within a dataset ({\it i.e.,} spectral features), and therefore ignore information that is highly specific to a given sample, ({\it i.e.,} the noise). Lastly, to facilitate convergence and training, we use the reconstructed spectra as our \YST{input} samples for the critic networks (see Equation ~\ref{eq:discrim_loss}). Therefore, the transfer of spectra from the synthetic to the observed domain is implicitly primed to produce noiseless spectra. Being able to learn from noisy spectra is particularly useful as we can train with the bulk of the survey data instead of restricting to only a small subset of high S/N spectra.

Since the network incorporates synthetic data in the training and is applied to real observations, \csn can also be regarded as an auto-calibration of the synthetic models. However, unlike standard calibration methods that are often based on a few standard stars \citep[{\it e.g.}, the Sun and Arcturus,][]{shetrone2015}, we effectively calibrate based on all of the stars in the dataset. The critical insight here is that spectra are fundamentally low-dimensional objects that lie on a manifold once transferred to an abstract latent-space. The latent space corresponds to the astrophysical properties of stars and has a finite and small number of degrees of freedom. Benefiting from a large amount of unlabelled data, we apply machine learning to discover this manifold that both synthetic and observed spectra lie on. By enforcing the commonality of the manifold, auto-calibration is attained. This is a drastically different philosophy to calibrate spectral models -- a calibration that relies on the redundancy in large datasets rather than standard ``ground truths.''  

As demonstrated in this study, our new insight can lead to a better calibration of spectral models than standard calibration techniques. This is perhaps not surprising because the standard calibration focuses only on a few stars, which span a limited range of stellar labels. Consequently, calibrating with relatively hot stars like the Sun and Arcturus requires extrapolation when considering cooler stars like the M-giants (see Figure~\ref{fig7} and Figure~\ref{fig9}). In contrast, \csn calibrates the models using all of the available stars, which span the entire stellar parameter space of interest. Furthermore, since the network learns from the existing data, it can capture the variations in the instrumental and observational effects. This allows the method to correct for things that are challenging to model with pre-existing methods ({\it e.g.}, the line spread function variation), alleviating a key roadblock when comparing models to observations.

\obriain{Finally, while we adopted stellar spectra as a case study, we emphasize that unsupervised domain adaptation could be broadly applied in astronomy as well as other fields. For instance, our method may be suitable for studies where computationally expensive simulations could be emulated, and large data sets exist to learn the corrections. It can be especially useful when, while the initial physics is simple, accurate modeling is infeasible due to the complexity of modeling subtle variations. For example, improved modeling of spectral energy distributions of stellar populations could be accomplished with \csnn. As another example, relating N-body simulations of dark matter distributions to large-scale galaxy data is an ideal candidate. For fields outside of astronomy, medical imaging provides interesting opportunities for \csn since the data is highly dependent on the particular medical device that is used for data acquisition. These instrumental effects could be corrected by either combining simulations and clinical data, or by learning the translation of data from one system to another. Last but not least, climate modelling is an area where efficiency could be improved by relating low resolution models to high resolution data -- a domain transfer problem where the difference between the domains is resolution.}

\subsection{Limitations and future implementations}
\label{sec:limitations}

While \csn has many attractive properties, the training of \csn can be delicate and may still require some fine-tuning. The difficulty mainly arises from the adversarial training setup. Specifically, adversarial training provides competition between the auto-encoder networks and the critic networks, making convergence not as straightforward as other machine learning tasks such as supervised regression. We note that the other proposed methods, including {\sc The Payne} \citep{ting2019}, {\sc The Cannon} \citep{ness2015cannon}, AstroNN \citep{leung2019}, and StarNet \citep{bialek2020} are examples of supervised regression, and therefore are technically much easier to train. 

To train \csnn, we have extensively explored numerous architectural choices and training details before deciding upon the final model. For instance, removing the adversarial losses leads to a smoother convergence, but without it, the latent representations are no longer shared, causing cross-domain translations to fail. Nevertheless, adversarial training is an active field of research in machine learning, with many new ideas proposed~\citep{Miyato18,Deshpande18,Donahue19} since the original \texttt{UNIT} paper was published. Consequently, we expect that some of these advancements would benefit and stabilize the training of future versions of \csnn.

Besides the difficulty in training in an adversarial setting, we have made a few simplifying assumptions regarding the training spectra data sets. Firstly, we assumed that the uncertainties provided by APOGEE are uncorrelated and accurately determined. How a mis-characterization of the noise may skew the training is yet to be studied.

\obriain{Additionally, the labels from both domains in this study span the same ranges. Early experiments suggest that having the two domains span a similar label space is an essential aspect of \csnn. For a new survey, we might not know beforehand the range of the stellar labels. How to resolve this problem, as well as efficient sampling of the $\geq 25$ parameter space dimensions, requires a more thorough investigation. While in principle, the training of \csn does not require paired samples in the two domains -- and is therefore unsupervised -- here we accelerated the training by matching distributions from APOGEE-Payne previously estimated stellar parameters, and generating synthetic spectra by drawing from these distributions. The use of the APOGEE-Payne distribution to generate the synthetic domain implies that we are not entirely blind when training the domain adaptation process. Perhaps \textit{weakly} supervised is a more appropriate description of the framework.}

Furthermore, since \csn is trained on the ``bulk'' of the observed data, the network might not extrapolate well for out-of-distribution samples and exotic stellar objects. To mitigate the potential impact of outliers during training, we eliminated APOGEE spectra by only considering spectra that produced a decent fit in the original APOGEE-Payne (reduced $\chi_R^2 < 50$). We do not expect a small number of odd samples to dominate the training, however, when adopting this method for other applications, it may be useful to remove outliers from the training process ({\it e.g.,} through ``simpler'' unsupervised clustering methods).

Recall that the domain adaptation process of \csn conducts a shared abstraction of both domains. The abstraction is then traced back to the stellar labels via a physics surrogate emulator. As a result, \csn auto-calibrates spectra primarily based on the original models, with which the emulator is trained. As demonstrated in Figure~\ref{fig10}, some systematics can persist even after the auto-calibration. For example, the $T_{\rm eff}$ and $\log g$ of the red clump stars are still systematically lower than the isochrones, similar to the results from the original Kurucz models. Additionally, the stellar parameters for the coolest dwarfs remain problematic. These results illustrate that the auto-calibration of \csn only works to a certain extent and may not correct for significant ``zero-point'' biases in the models. Including external physical priors ({\it e.g.,} an isochrone prior) could constrain the zero-point biases.

\obriain{With regards to the case study presented in \secref{results_mocknewlines} -- where the flux responses of \csn were investigated to identify missing information in the synthetic modelling -- the application to real observations can present new challenges. For instance, the difference in the flux responses will no longer be strictly due to missing spectral lines; inaccurate oscillator strengths or damping, for example, can cause discrepancies between the flux responses. This is by design and exemplifies how \csn is correcting for these effects, but this adds more complexity to the analysis. However, since it is impossible to synthesize all possible unknowns, in practice, the flux responses should be examined on a case-by-case basis and domain knowledge should be utilized.}

As a final remark, \csn is a project that began 2 years ago. During the development of \csnn, multiple unsupervised domain adaptation methods have emerged, with potentials to alleviate some of the mentioned caveats. In particular, the UNIT method has been adapted to multiple domain adaptation \citep{huang2018multimodal}, and to smaller sample requirements \citep{liu2019few}. Normalizing flows for domain adaptation have also shown to enforce cycle-consistency and provide more tractable gradients and likelihoods by construction \citep{grover2020alignflow}.

\subsection{Open source}
\label{sec:github}

For future reproducibility and to assist with applications to other projects, we have made the code for \csn publicly available on GitHub\footnote{ \href{https://github.com/teaghan/Cycle_SN}{https://github.com/teaghan/Cycle\_SN}}, along with in-depth explanations on the code itself and the training details. As noted in \Sec{limitations}, training \csn can be sensitive to the choices of the network architecture and training hyper-parameters. Therefore, we emphasize that the GitHub aims to only serve as a starting point for other applications.

%% file: 6_conclusion.tex
\section{Conclusion}
\label{sec:conclusion}

Maximally extracting information from stellar spectra requires that we have perfect knowledge of spectral synthesis, an idea that has remained elusive despite decades of studies. In this paper, we present a new methodology, \csnn, to tackle this problem.  \csn adopts ideas from Domain Adaptation in Machine Learning to mitigate model systematics. Our results are summarized below:

\begin{enumerate}
    \item \csn auto-calibrates for deficiencies in spectral modeling, and develops a common abstraction of both synthetic and observed data through an unsupervised network. This abstraction is related to physical stellar labels via a physics surrogate emulator network.
    
    \item \csn can build data-driven models via a set of unlabelled training spectra, without knowing the stellar labels of the training spectra a priori. 
    
    \item Through the use of a split latent space, \csn can distinguish the actual astrophysical information from the spectral variations due to instrumental and observational factors. This reduces the reliance of the modeling accuracy on instrumental factors, such as the fiber-dependent line spread functions and telluric feature modelling. 
    
    \item By fitting the APOGEE spectra, we demonstrated that the auto-calibrated models produced by \csn yield more precise stellar parameters than the original model, even without adopting any spectroscopic mask or spectral windows. 
    
    \item \csn can understand stellar astrophysics and uncover unknown spectral features. Testing on a mock dataset, we illustrated that \csn can recover the missing features in the Kurucz models and associate the missing features to the correct corresponding elements.
\end{enumerate}

The philosophy of auto-calibrating models with domain adaptation, exemplified by \csnn, is generic and can be applied to many other fields. Our results provide an entirely new path to extract information from large unlabelled datasets; harnessing advancements in Machine Learning to redefine what big data astronomy can mean for stellar spectroscopy and beyond.

%% file: 7_appendix.tex
\section{\csn Architecture}
\label{sec:appendix_architecture}

\input{tab2}

We summarize the architecture of each of the sub-networks in Table~\ref{table:network_architecture}. Each row within the table shows a subsequent network layer, as well as the activation function (LeakyReLU, Sigmoid, InstanceNorm) applied after the layer operation. The CONV layers are the standard 1D-convolutional layers; the DCONV layers are de-convolutional (or transposed convolutional) layers; and FC denotes the fully-connected layers. For each layer, N, K, and S represent, respectively, the number of filters (or nodes), the size of the filters (or kernel size), and the stride-length of the convolutional operations.

\obriain{Furthermore, instead of using the standard 7214 pixels from the ASPCAP reduction, we discard 47 pixels in the red chip to easily accommodate symmetrical down- and up-sampling in the convolutional networks. Thus, each spectrum consists of 7167 pixels. When processed through either $\Es$ or $\Eo$, the intermediate representation is $64\times 447$ dimensional. This intermediate representation is downsampled by $\Esh$ to produce the shared latent representation, which is $25\times 24$ dimensional. Lastly, $\Esp$ downsamples the intermediate representation to the split latent space, which either has the shape $4\times 24$ (\secref{results_domaintransfer}) or $1\times 24$ (\secref{results_mocknewlines}).}

On this note, the only difference between the architectures used in Sections \ref{sec:results_domaintransfer} and \ref{sec:results_mocknewlines} is that the split latent-space has four filters in the former and one filter in the latter. Note that, since Section \ref{sec:results_mocknewlines} uses a mock dataset for the observed domain, there is no other observation-specific variables. Therefore, in principle, there is no need for a split latent-space. However, to keep the two architectures relatively consistent, we simply used fewer filters.

\section{\csn Training Details}
\label{sec:appendix_training_details}

\input{fig15}

\csn is trained by optimizing the loss outlined in Equation \ref{eq:total_loss}. An essential aspect of having the network converge correctly is determining the correct combination of the $\lambda$ values in this equation, which control the influence of each loss term. Through various empirical tests, we determined that the values $\lambda_{synth}=3$, $\lambda_{obs}=9$, and $\lambda_{adv}=1$ lead to satisfactory performance.

As discussed in \Sec{adversarial}, training is done iteratively; one iteration of training the auto-encoders is computed -- including the adversarial loss for the transfer mappings -- followed by an iteration of training the critic networks. Each iteration used a batch of 8 spectra from each domain. If we consider a single batch iteration as one optimization step for both processes, the network was trained for 350,000 batch iterations. The training was done with the Adam optimizer using a learning rate of 0.0001. Furthermore, this learning rate was decreased by a factor of 0.7 at 50k, 100k, 150k, and 200k batch iterations. \obriain{\Fig{fig15} shows the training progress for the case study presented in \secref{results_domaintransfer}.}

%% file: tab2.tex
\begin{table}
\caption{A summary of the sub-network architectures.
\label{table:network_architecture}}
\begin{tabular}{cc}
\hline
Layer & $E_{synth}$ \&  $E_{obs}$   \\ \hline
1     & CONV-(N32,K7,S4), LeakyReLU       \\
2     & CONV-(N64,K7,S4), LeakyReLU       \\
\hline
Layer & $E_{sh}$  \\ 
\hline
1     & CONV-(N128,K7,S4), LeakyReLU     \\ 
2     & CONV-(N256,K7,S2), LeakyReLU      \\
3     & CONV-(N512,K7,S2), LeakyReLU     \\
4     & CONV-(N25,K1,S1), InstanceNorm    \\ 
\hline
Layer & $E_{sp}$  \\ 
\hline
1     & CONV-(N32,K7,S4), LeakyReLU    \\
2     & CONV-(N32,K7,S2), LeakyReLU    \\
3     & CONV-(N32,K7,S2), LeakyReLU    \\
4     & CONV-(N4(1),K1,S1), InstanceNorm    \\ 
\hline
Layer & $D_{sp}$    \\ 
\hline
1     & DCONV-(N32,K7,S2), LeakyReLU     \\
2     & DCONV-(N32,K7,S2), LeakyReLU          \\
3     & DCONV-(N32,K7,S4), LeakyReLU          \\
\hline
Layer & $D_{sh}$    \\ 
\hline
1     & DCONV-(N512,K7,S2), LeakyReLU     \\
2     & DCONV-(N256,K7,S2), LeakyReLU          \\
3     & DCONV-(N128,K7,S4), LeakyReLU          \\
\hline
Layer & $D_{synth}$ \&  $D_{obs}$   \\ 
\hline
1     & DCONV-(N64,K7,S4), LeakyReLU          \\
2     & DCONV-(N32,K7,S4), LeakyReLU          \\
3     & CONV-(N1,K1,S1)    \\ 
\hline
Layer & $C_{synth}$ \&  $C_{obs}$     \\ 
\hline
1a     & CONV-(N16,K7,S4), LeakyReLU         \\
2a     & CONV-(N32,K7,S4), LeakyReLU        \\
3a     & CONV-(N64,K7,S4), LeakyReLU      \\
4a     & CONV-(N128,K7,S4), LeakyReLU        \\
5a     & CONV-(N256,K7,S4), LeakyReLU      \\
1b     & CONV-(N32,K1,S1), LeakyReLU        \\
2b     & CONV-(N64,K1,S1), LeakyReLU      \\
3b     & CONV-(N128,K1,S1), LeakyReLU        \\
4b     & CONV-(N256,K1,S1), LeakyReLU      \\
5b     & CONV-(N512,K1,S1), LeakyReLU      \\
6     & FC-(N1), Sigmoid  \\
\hline
\end{tabular}
\end{table}

%% file: fig15.tex
\begin{figure}
\includegraphics[width=\linewidth]{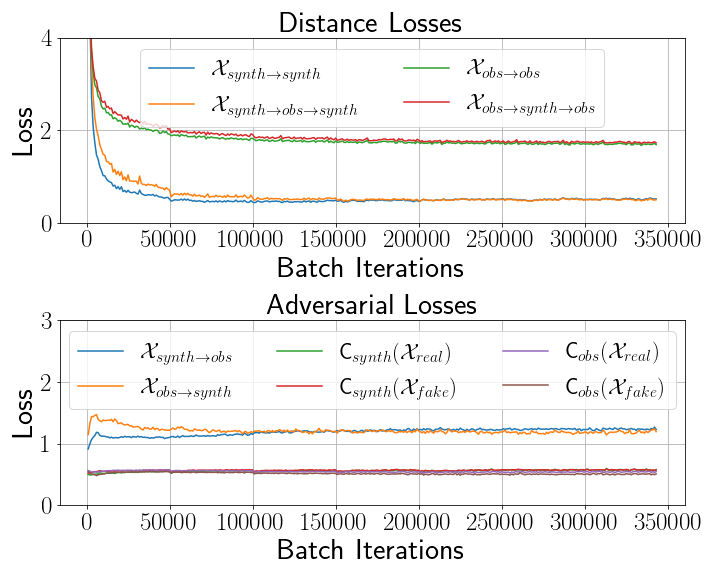}
\caption{\obriain{The training progress for the case study outlined in \secref{results_domaintransfer}. The top panel reflects the distance losses in the both the synthetic and observed domains for the training reconstructions and cycle-reconstructions. The bottom panel shows the evaluation of the adversarial losses on the training set for both the generative processes (\Xonetwo\ and \Xtwoone) and the critics (represented simply as $C_{synth/obs}(\mathcal{X}_{true/fake})$).}
\label{fig15}}
\end{figure}

%% file: manuscript.bbl
\begin{thebibliography}{}
\expandafter\ifx\csname natexlab\endcsname\relax\def\natexlab#1{#1}\fi
\providecommand{\url}[1]{\href{#1}{#1}}
\providecommand{\dodoi}[1]{}
\providecommand{\doeprint}[1]{\href{http://ascl.net/#1}{\nolinkurl{http://ascl.net/#1}}}
\providecommand{\doarXiv}[1]{\href{https://arxiv.org/abs/#1}{\nolinkurl{https://arxiv.org/abs/#1}}}

\bibitem[{{Amarsi}(2015)}]{amarsi2015}
{Amarsi}, A.~M. 2015,
  \href{http://dx.doi.org/10.1093/mnras/stv1392}{\color{magenta}\mnras},
  \href{https://ui.adsabs.harvard.edu/abs/2015MNRAS.452.1612A}{\color{cyan}452},
  1612

\bibitem[{{Amarsi} {et~al.}(2016){Amarsi}, {Asplund}, {Collet}, \&
  {Leenaarts}}]{amarsi2016}
{Amarsi}, A.~M., {Asplund}, M., {Collet}, R., \& {Leenaarts}, J. 2016,
  \href{http://dx.doi.org/10.1093/mnras/stv2608}{\color{magenta}\mnras},
  \href{https://ui.adsabs.harvard.edu/abs/2016MNRAS.455.3735A}{\color{cyan}455},
  3735

\bibitem[{{Aoki} {et~al.}(2013){Aoki}, {Beers}, {Lee}, {Honda}, {Ito},
  {Takada-Hidai}, {Frebel}, {Suda}, {Fujimoto}, {Carollo}, \&
  {Sivarani}}]{aoki2013hires}
{Aoki}, W., {Beers}, T.~C., {Lee}, Y.~S., {et~al.} 2013, \aj, 145, 13

\bibitem[{{Ballester} {et~al.}(2000){Ballester}, {Modigliani}, {Boitquin},
  {Cristiani}, {Hanuschik}, {Kaufer}, \& {Wolf}}]{ballester2000}
{Ballester}, P., {Modigliani}, A., {Boitquin}, O., {et~al.} 2000, The
  Messenger,
  \href{https://ui.adsabs.harvard.edu/abs/2000Msngr.101...31B}{\color{cyan}101},
  31

\bibitem[{Bialek {et~al.}(2020)Bialek, Fabbro, Venn, Kumar, O'Briain, \&
  Yi}]{bialek2020}
Bialek, S., Fabbro, S., Venn, K.~A., {et~al.} 2020, in prep

\bibitem[{{Buder} {et~al.}(2018){Buder}, {Asplund}, {Duong}, {Kos}, {Lind},
  {Ness}, {Sharma}, {Bland -Hawthorn}, {Casey}, {de Silva}, {D'Orazi},
  {Freeman}, {Lewis}, {Lin}, {Martell}, {Schlesinger}, {Simpson}, {Zucker},
  {Zwitter}, {Amarsi}, {Anguiano}, {Carollo}, {Casagrande}, {{\v{C}}otar},
  {Cottrell}, {da Costa}, {Gao}, {Hayden}, {Horner}, {Ireland}, {Kafle},
  {Munari}, {Nataf}, {Nordlander}, {Stello}, {Ting}, {Traven}, {Watson},
  {Wittenmyer}, {Wyse}, {Yong}, {Zinn}, {{\v{Z}}erjal}, \& {Galah
  Collaboration}}]{buder2018}
{Buder}, S., {Asplund}, M., {Duong}, L., {et~al.} 2018,
  \href{http://dx.doi.org/10.1093/mnras/sty1281}{\color{magenta}\mnras},
  \href{https://ui.adsabs.harvard.edu/abs/2018MNRAS.478.4513B}{\color{cyan}478},
  4513

\bibitem[{{Coelho} {et~al.}(2020){Coelho}, {Bruzual}, \&
  {Charlot}}]{coelho2020}
{Coelho}, P. R.~T., {Bruzual}, G., \& {Charlot}, S. 2020,
  \href{http://dx.doi.org/10.1093/mnras/stz3023}{\color{magenta}\mnras},
  \href{https://ui.adsabs.harvard.edu/abs/2020MNRAS.491.2025C}{\color{cyan}491},
  2025

\bibitem[{{Dalton} {et~al.}(2014){Dalton}, {Trager}, {Abrams}, {Bonifacio},
  {L{\'o}pez Aguerri}, {Middleton}, {Benn}, {Dee}, {Say{\`e}de}, {Lewis},
  {Pragt}, {Pico}, {Walton}, {Rey}, {Allende Prieto}, {Pe{\~n}ate}, {Lhome},
  {Ag{\'o}cs}, {Alonso}, {Terrett}, {Brock}, {Gilbert}, {Ridings}, {Guinouard},
  {Verheijen}, {Tosh}, {Rogers}, {Steele}, {Stuik}, {Tromp}, {Jasko}, {Kragt},
  {Lesman}, {Mottram}, {Bates}, {Gribbin}, {Fernand o Rodriguez}, {Delgado},
  {Martin}, {Cano}, {Navarro}, {Irwin}, {Lewis}, {Gonzalez Solares},
  {O'Mahony}, {Bianco}, {Zurita}, {ter Horst}, {Molinari}, {Lodi}, {Guerra},
  {Vallenari}, \& {Baruffolo}}]{dalton2014}
{Dalton}, G., {Trager}, S., {Abrams}, D.~C., {et~al.} 2014, in
  \href{http://dx.doi.org/10.1117/12.2055132}{\color{magenta}Society of
  Photo-Optical Instrumentation Engineers (SPIE) Conference Series}, Vol.
  \href{https://ui.adsabs.harvard.edu/abs/2014SPIE.9147E..0LD}{\color{cyan}9147},
  \procspie, 91470L

\bibitem[{Deshpande {et~al.}(2018)Deshpande, Zhang, \& Schwing}]{Deshpande18}
Deshpande, I., Zhang, Z., \& Schwing, A.~G. 2018, in Conference on Computer
  Vision and Pattern Recognition

\bibitem[{Donahue \& Simonyan(2019)}]{Donahue19}
Donahue, J., \& Simonyan, K. 2019, in Advances in Neural Information Processing
  Systems, 10542

\bibitem[{{Fabbro} {et~al.}(2018){Fabbro}, {Venn}, {O'Briain}, {Bialek},
  {Kielty}, {Jahandar}, \& {Monty}}]{fabbro2018}
{Fabbro}, S., {Venn}, K.~A., {O'Briain}, T., {et~al.} 2018,
  \href{http://dx.doi.org/10.1093/mnras/stx3298}{\color{magenta}\mnras},
  \href{https://ui.adsabs.harvard.edu/abs/2018MNRAS.475.2978F}{\color{cyan}475},
  2978

\bibitem[{{Gilmore} {et~al.}(2012){Gilmore}, {Randich}, {Asplund}, {Binney},
  {Bonifacio}, {Drew}, {Feltzing}, {Ferguson}, {Jeffries}, {Micela}, \&
  et~al.}]{gilmore2012gaiaeso}
{Gilmore}, G., {Randich}, S., {Asplund}, M., {et~al.} 2012, The Messenger, 147,
  25

\bibitem[{Gonzalez{-}Garcia {et~al.}(2018)Gonzalez{-}Garcia, van~de Weijer, \&
  Bengio}]{gonzalez2018}
Gonzalez{-}Garcia, A., van~de Weijer, J., \& Bengio, Y. 2018, CoRR,
  abs/1805.09730.
\newblock \doarXiv{1805.09730}

\bibitem[{Goodfellow {et~al.}(2014)Goodfellow, Pouget-Abadie, Mirza, Xu,
  Warde-Farley, Ozair, Courville, \& Bengio}]{goodfellow2014}
Goodfellow, I.~J., Pouget-Abadie, J., Mirza, M., {et~al.} 2014, NIPS, 2672

\bibitem[{Grover {et~al.}(2020)Grover, Chute, Shu, Cao, \&
  Ermon}]{grover2020alignflow}
Grover, A., Chute, C., Shu, R., Cao, Z., \& Ermon, S. 2020, in AAAI, 4028

\bibitem[{{Guiglion} {et~al.}(2020){Guiglion}, {Matijevic}, {Queiroz},
  {Valentini}, {Steinmetz}, {Chiappini}, {Grebel}, {McMillan}, {Kordopatis},
  {Kunder}, {Zwitter}, {Khalatyan}, {Anders}, {Enke}, {Minchev}, {Monari},
  {Wyse}, {Bienayme}, {Bland-Hawthorn}, {Gibson}, {Navarro}, {Parker}, {Reid},
  {Seabroke}, \& {Siebert}}]{guiglion2020}
{Guiglion}, G., {Matijevic}, G., {Queiroz}, A.~B.~A., {et~al.} 2020, arXiv
  e-prints, arXiv:2004.12666.
\newblock \doarXiv{2004.12666}

\bibitem[{{Holtzman} {et~al.}(2018){Holtzman}, {Hasselquist}, {Shetrone},
  {Cunha}, {Allende Prieto}, {Anguiano}, {Bizyaev}, {Bovy}, {Casey},
  {Edvardsson}, {Johnson}, {J{\"o}nsson}, {Meszaros}, {Smith}, {Sobeck},
  {Zamora}, {Chojnowski}, {Fernandez-Trincado}, {Garcia-Hernandez}, {Majewski},
  {Pinsonneault}, {Souto}, {Stringfellow}, {Tayar}, {Troup}, \&
  {Zasowski}}]{holtzman2018}
{Holtzman}, J.~A., {Hasselquist}, S., {Shetrone}, M., {et~al.} 2018,
  \href{http://dx.doi.org/10.3847/1538-3881/aad4f9}{\color{magenta}\aj},
  \href{https://ui.adsabs.harvard.edu/abs/2018AJ....156..125H}{\color{cyan}156},
  125

\bibitem[{Huang {et~al.}(2018)Huang, Liu, Belongie, \&
  Kautz}]{huang2018multimodal}
Huang, X., Liu, M.-Y., Belongie, S., \& Kautz, J. 2018, in Proceedings of the
  European Conference on Computer Vision (ECCV), 172

\bibitem[{{Jahandar} {et~al.}(2017){Jahandar}, {Venn}, {Shetrone}, {Irwin},
  {Bovy}, {Sakari}, {Kielty}, {Digby}, \& {Frinchaboy}}]{jahandar2017}
{Jahandar}, F., {Venn}, K.~A., {Shetrone}, M.~D., {et~al.} 2017,
  \href{http://dx.doi.org/10.1093/mnras/stx1592}{\color{magenta}\mnras},
  \href{https://ui.adsabs.harvard.edu/abs/2017MNRAS.470.4782J}{\color{cyan}470},
  4782

\bibitem[{{Kovalev} {et~al.}(2019){Kovalev}, {Bergemann}, {Ting}, \&
  {Rix}}]{kovalev2019}
{Kovalev}, M., {Bergemann}, M., {Ting}, Y.-S., \& {Rix}, H.-W. 2019,
  \href{http://dx.doi.org/10.1051/0004-6361/201935861}{\color{magenta}\aap},
  \href{https://ui.adsabs.harvard.edu/abs/2019A&A...628A..54K}{\color{cyan}628},
  A54

\bibitem[{{Kurucz}(1970)}]{kurucz1970}
{Kurucz}, R.~L. 1970, SAO Special Report,
  \href{http://adsabs.harvard.edu/abs/1970SAOSR.309.....K}{\color{cyan}309, 291
  pp}

\bibitem[{{Kurucz}(1993{\natexlab{a}})}]{kurucz1993}
---. 1993{\natexlab{a}}, {SYNTHE spectrum synthesis programs and line data}

\bibitem[{{Kurucz}(1993{\natexlab{b}})}]{kur93}
---. 1993{\natexlab{b}}, {SYNTHE spectrum synthesis programs and line data},
  ed. {Kurucz, R.~L.}

\bibitem[{{Kurucz}(2005{\natexlab{a}})}]{kurucz2005}
---. 2005{\natexlab{a}}, Memorie della Societa Astronomica Italiana
  Supplementi,
  \href{https://ui.adsabs.harvard.edu/abs/2005MSAIS...8...14K}{\color{cyan}8},
  14

\bibitem[{{Kurucz}(2005{\natexlab{b}})}]{kur05}
---. 2005{\natexlab{b}}, Memorie della Societa Astronomica Italiana
  Supplementi,
  \href{http://adsabs.harvard.edu/abs/2005MSAIS...8...14K}{\color{cyan}8}, 14

\bibitem[{{Kurucz}(2013)}]{kur13}
---. 2013, {ATLAS12: Opacity sampling model atmosphere program}, Astrophysics
  Source Code Library.
\newblock \doeprint{1303.024}

\bibitem[{{Kurucz} \& {Avrett}(1981{\natexlab{a}})}]{kurucz1981}
{Kurucz}, R.~L., \& {Avrett}, E.~H. 1981{\natexlab{a}}, SAO Special Report,
  \href{http://adsabs.harvard.edu/abs/1981SAOSR.391.....K}{\color{cyan}391, 139
  pp}

\bibitem[{{Kurucz} \& {Avrett}(1981{\natexlab{b}})}]{kur81}
---. 1981{\natexlab{b}}, SAO Special Report,
  \href{http://adsabs.harvard.edu/abs/1981SAOSR.391.....K}{\color{cyan}391, 139
  pp}

\bibitem[{{Leung} \& {Bovy}(2019)}]{leung2019}
{Leung}, H.~W., \& {Bovy}, J. 2019,
  \href{http://dx.doi.org/10.1093/mnras/sty3217}{\color{magenta}\mnras},
  \href{https://ui.adsabs.harvard.edu/abs/2019MNRAS.483.3255L}{\color{cyan}483},
  3255

\bibitem[{Liu {et~al.}(2017)Liu, Breuel, \& Kautz}]{liu2017}
Liu, M.-Y., Breuel, T., \& Kautz, J. 2017, NIPS

\bibitem[{Liu {et~al.}(2019)Liu, Huang, Mallya, Karras, Aila, Lehtinen, \&
  Kautz}]{liu2019few}
Liu, M.-Y., Huang, X., Mallya, A., {et~al.} 2019, in Proceedings of the IEEE
  International Conference on Computer Vision, 10551

\bibitem[{Maaten \& Hinton(2008)}]{maaten2008visualizing}
Maaten, L. v.~d., \& Hinton, G. 2008, Journal of Machine Learning Research, 9,
  2579

\bibitem[{{Martioli} {et~al.}(2012){Martioli}, {Teeple}, {Manset}, {Devost},
  {Withington}, {Venne}, \& {Tannock}}]{opera2012}
{Martioli}, E., {Teeple}, D., {Manset}, N., {et~al.} 2012, in
  \href{http://dx.doi.org/10.1117/12.926627}{\color{magenta}Society of
  Photo-Optical Instrumentation Engineers (SPIE) Conference Series}, Vol.
  \href{https://ui.adsabs.harvard.edu/abs/2012SPIE.8451E..2BM}{\color{cyan}8451},
  \procspie, 84512B

\bibitem[{Miyato {et~al.}(2018)Miyato, Kataoka, Koyama, \& Yoshida}]{Miyato18}
Miyato, T., Kataoka, T., Koyama, M., \& Yoshida, Y. 2018, in International
  Conference on Learning Representations

\bibitem[{Ness {et~al.}(2015)Ness, Hogg, Rix, Ho, \& Zasowski}]{ness2015cannon}
Ness, M., Hogg, D.~W., Rix, H.-W., Ho, A.~Y., \& Zasowski, G. 2015, \apj, 808,
  16

\bibitem[{Prugniel {et~al.}(2011)Prugniel, Vauglin, \& Koleva}]{Prugniel2011}
Prugniel, P., Vauglin, I., \& Koleva, M. 2011,
  \href{http://dx.doi.org/10.1051/0004-6361/201116769}{\color{magenta}Astronomy
  \& Astrophysics}, 531, A165

\bibitem[{Sharma {et~al.}(2015)Sharma, Prugniel, \& Singh}]{Sharma2015}
Sharma, K., Prugniel, P., \& Singh, H.~P. 2015,
  \href{http://dx.doi.org/10.1051/0004-6361/201526111}{\color{magenta}Astronomy
  \& Astrophysics}, 585, A64

\bibitem[{Shetrone {et~al.}(2015)Shetrone, Bizyaev, Lawler, Allende, A., Smith,
  Cunha, Holtzman, Garcia, Meszaros, Sobeck, Zamora, Garcia-Hernandez, Souto,
  Chojnowski, Koesterke, Majewski, \& Zasowski}]{shetrone2015}
Shetrone, M., Bizyaev, D., Lawler, J., {et~al.} 2015, The American Astronomical
  Society, 221

\bibitem[{{Sneden} {et~al.}(2008){Sneden}, {Cowan}, \& {Gallino}}]{sneden2008}
{Sneden}, C., {Cowan}, J.~J., \& {Gallino}, R. 2008,
  \href{http://dx.doi.org/10.1146/annurev.astro.46.060407.145207}{\color{magenta}\araa},
  \href{https://ui.adsabs.harvard.edu/abs/2008ARA&A..46..241S}{\color{cyan}46},
  241

\bibitem[{{Ting} {et~al.}(2018){Ting}, {Conroy}, {Rix}, \&
  {Asplund}}]{ting2018}
{Ting}, Y.-S., {Conroy}, C., {Rix}, H.-W., \& {Asplund}, M. 2018,
  \href{http://dx.doi.org/10.3847/1538-4357/aac6c9}{\color{magenta}\apj},
  \href{https://ui.adsabs.harvard.edu/abs/2018ApJ...860..159T}{\color{cyan}860},
  159

\bibitem[{{Ting} {et~al.}(2019){Ting}, {Conroy}, {Rix}, \&
  {Cargile}}]{ting2019}
{Ting}, Y.-S., {Conroy}, C., {Rix}, H.-W., \& {Cargile}, P. 2019,
  \href{http://dx.doi.org/10.3847/1538-4357/ab2331}{\color{magenta}\apj},
  \href{https://ui.adsabs.harvard.edu/abs/2019ApJ...879...69T}{\color{cyan}879},
  69

\bibitem[{{Ting} {et~al.}(2017{\natexlab{a}}){Ting}, {Rix}, {Conroy}, {Ho}, \&
  {Lin}}]{ting2017}
{Ting}, Y.-S., {Rix}, H.-W., {Conroy}, C., {Ho}, A. Y.~Q., \& {Lin}, J.
  2017{\natexlab{a}},
  \href{http://dx.doi.org/10.3847/2041-8213/aa921c}{\color{magenta}\apjl},
  \href{https://ui.adsabs.harvard.edu/abs/2017ApJ...849L...9T}{\color{cyan}849},
  L9

\bibitem[{{Ting} {et~al.}(2017{\natexlab{b}}){Ting}, {Rix}, {Conroy}, {Ho}, \&
  {Lin}}]{Ting2017b}
---. 2017{\natexlab{b}},
  \href{http://dx.doi.org/10.3847/2041-8213/aa921c}{\color{magenta}\apjl},
  \href{https://ui.adsabs.harvard.edu/abs/2017ApJ...849L...9T}{\color{cyan}849},
  L9

\bibitem[{{Venn} {et~al.}(2020){Venn}, {Kielty}, {Sestito}, {Starkenburg},
  {Martin}, {Aguado}, {Arentsen}, {Bonifacio}, {Caffau}, {Hill}, {Jablonka},
  {Lardo}, {Mashonkina}, {Navarro}, {Sneden}, {Thomas}, {Youakim},
  {Gonz{\'a}lez-Hern{\'a}ndez}, {S{\'a}nchez Janssen}, {Carlberg}, \&
  {Malhan}}]{venn2020}
{Venn}, K.~A., {Kielty}, C.~L., {Sestito}, F., {et~al.} 2020,
  \href{http://dx.doi.org/10.1093/mnras/stz3546}{\color{magenta}\mnras},
  \href{https://ui.adsabs.harvard.edu/abs/2020MNRAS.492.3241V}{\color{cyan}492},
  3241

\bibitem[{{Xiang} {et~al.}(2019){Xiang}, {Ting}, {Rix}, {Sand ford}, {Buder},
  {Lind}, {Liu}, {Shi}, \& {Zhang}}]{Xiang2019}
{Xiang}, M., {Ting}, Y.-S., {Rix}, H.-W., {et~al.} 2019,
  \href{http://dx.doi.org/10.3847/1538-4365/ab5364}{\color{magenta}\apjs},
  \href{https://ui.adsabs.harvard.edu/abs/2019ApJS..245...34X}{\color{cyan}245},
  34

\bibitem[{Yanny {et~al.}(2009)Yanny, Rockosi, Newberg, Knapp, Adelman-McCarthy,
  Alcorn, Allam, Prieto, An, Anderson, {et~al.}}]{yanny2009segue}
Yanny, B., Rockosi, C., Newberg, H.~J., {et~al.} 2009, The Astronomical
  Journal, 137, 4377

\bibitem[{{Zhang} {et~al.}(2019){Zhang}, {Zhao}, {Yang}, {Wang}, \&
  {Zuo}}]{zhang2019}
{Zhang}, X., {Zhao}, G., {Yang}, C.~Q., {Wang}, Q.~X., \& {Zuo}, W.~B. 2019,
  \href{http://dx.doi.org/10.1088/1538-3873/ab2687}{\color{magenta}\pasp},
  \href{https://ui.adsabs.harvard.edu/abs/2019PASP..131i4202Z}{\color{cyan}131},
  094202

\bibitem[{Zhu {et~al.}(2017)Zhu, Park, Isola, \& Efros}]{zhu2017unpaired}
Zhu, J.-Y., Park, T., Isola, P., \& Efros, A.~A. 2017, in IEEE International
  Conference on Computer Vision, 2223

\end{thebibliography}
